\documentclass[10pt,aps,prb,twocolumn,superscriptaddress]{revtex4-2}
\usepackage[utf8]{inputenc}
\usepackage{graphicx, color}
\usepackage{xcolor}
\usepackage{amsmath}
\usepackage{mathtools}
\usepackage{amssymb}
\usepackage[version=4]{mhchem}
\usepackage{bm}
\usepackage{siunitx}
\usepackage{gensymb}
\usepackage{multirow}
\usepackage{textcomp}
\usepackage{float}
\usepackage{array}
\usepackage{longtable}
\usepackage{tabularray}
\usepackage{booktabs}
\usepackage{url}
\usepackage{nameref}
\usepackage{hyperref}
\usepackage{xspace}
\usepackage{tikz}
\usepackage[toc,page]{appendix}
\definecolor{myblue}{rgb}{0,0,180}
\definecolor{link}{rgb}{0.1,0.1,0.9}
\hypersetup{colorlinks=true,linkcolor=link,citecolor=link,urlcolor=link,linktocpage}

\makeatletter
\g@addto@macro\bfseries{\boldmath}
\makeatother

\DeclareUnicodeCharacter{2212}{\textendash}

\begin{document}
	
	\preprint{APS/123-QED}
	
	\title{Controlling Quantum Materials by Growth: Thermodynamics, Kinetics, and Defect Engineering in Transition Metal Dichalcogenides}
	
	\author{Anzar Ali}
	\email{Anzar.Ali@udst.edu.qa}
	\affiliation{University of Doha for Science and Technology, P.O. Box 24449, Doha, Qatar}
	
	\author{Md Ezaz Hasan Khan}
	\affiliation{University of Doha for Science and Technology, P.O. Box 24449, Doha, Qatar}
	
	\author{Mahmoud Abdel-Hafiez}
	\email{mahmoudhafiez@gmail.com}
	\affiliation{University of Doha for Science and Technology, P.O. Box 24449, Doha, Qatar}
	\affiliation{Department of Applied Physics and Astronomy, University of Sharjah, P.O. Box 27272, Sharjah, United Arab Emirates}

	\begin{abstract}
		Transition metal dichalcogenides exhibit a wide range of semiconducting, metallic, correlated, and topological electronic states that arise from strong coupling between lattice structure, dimensionality, and electronic degrees of freedom. In these materials, crystal growth is not merely a preparative step but a thermodynamic boundary condition that establishes chemical potentials, defect populations, polytype stability, and access to metastable phases. As a result, synthesis determines the structural and defect landscape from which collective electronic behavior emerges.
		
		In this Review, we develop a unified thermodynamic and kinetic framework that connects growth conditions to phase stability, defect energetics, and microstructure. We examine how chemical potential constraints define stability windows, how supersaturation and mass-transport regimes govern nucleation and morphology, and how nonequilibrium pathways enable kinetic trapping and polymorph selection. Bulk and thin-film synthesis approaches, including chemical vapor transport, flux growth, physical vapor transport, solvent-assisted crystallization, chemical vapor deposition, and molecular beam epitaxy, are placed within a common thermodynamic and kinetic map to clarify how distinct growth regimes produce characteristic disorder profiles and structural phases.
		
		By explicitly linking synthesis variables to charge-density-wave order, superconductivity, band topology, and correlation effects, this Review establishes crystal growth as a central parameter in determining the effective electronic Hamiltonian realized in experiment. This perspective provides a physically grounded framework for improving reproducibility and guiding deterministic control of emergent quantum phases in layered materials.
	\end{abstract}

	
	\maketitle
	

\section{Introduction}
\label{Intro}

Transition metal dichalcogenides (TMDs) form a broad class of layered compounds with the chemical formula MX$_2$, where M is a transition metal and X is a chalcogen. Their crystal structure consists of covalently bonded X–M–X trilayers stacked through weak van der Waals interactions. This bonding hierarchy produces strong anisotropy in bonding, transport, and mechanical response~\cite{Manzeli_2017,Wilson_1975}. The layered structure allows exfoliation down to the monolayer limit and enables systematic investigation of dimensional crossover without chemical surface reconstruction. As a result, TMDs have become model systems for studying how lattice symmetry, reduced dimensionality, and electronic structure govern collective electronic behavior.

The electronic ground states realized in TMDs span semiconducting, metallic, correlated, and topological regimes. Semiconducting compounds such as 2H-MoS$_2$, WS$_2$, MoSe$_2$, and WSe$_2$ exhibit thickness-dependent band structures, including the indirect-to-direct band-gap transition in the monolayer limit~\cite{Mak_2010,Splendiani_2010,Manzeli_2017}. Metallic compounds such as 2H-NbSe$_2$ and 2H-TaS$_2$ display intertwined charge-density-wave (CDW) order and superconductivity~\cite{Wilson_1975,Moncton_1975}. In 1T-TiSe$_2$ and 1T-TaS$_2$, collective possible excitonic instability produce commensurate CDW phases and, in certain regimes, Mott-like insulating states at low temperature~\cite{DiSalvo_1976,Wilson_1975,Fazekas_1979}. Telluride systems including MoTe$_2$ and WTe$_2$ exhibit large nonsaturating magnetoresistance and symmetry-sensitive electronic structures in which subtle lattice distortions govern topological character~\cite{Ali_2014,Qian_2014,Soluyanov_2015}.

A defining characteristic of this materials family is the proximity of competing phases separated by small energy scales. In NbSe$_2$ and TaS$_2$, CDW order competes with superconductivity within a narrow thermodynamic window~\cite{Wilson_1975,Moncton_1975}. In TiSe$_2$, carrier concentration tunes the balance between semimetallic band overlap and excitonic instability~\cite{DiSalvo_1976}. In MoTe$_2$ and WTe$_2$, structural distortions of only a few percent modify inversion symmetry and alter band topology~\cite{Qian_2014,Soluyanov_2015}. Because the relevant free-energy differences are often on the order of tens of meV per formula unit, modest variations in stoichiometry, strain, or disorder can shift phase boundaries in measurable ways.

Polymorphism further expands the accessible phase space. For a fixed composition, distinct coordination geometries and stacking sequences, including 2H, 1T, 1T$'$, and T$_\mathrm{d}$ polytypes, may be stabilized~\cite{Chen_2022}. The relative stability of these phases is determined by the Gibbs free energy,
\begin{equation}
	G = H - TS,
\end{equation}
where enthalpic contributions arising from bonding and electronic structure compete with entropic contributions that increase with temperature. Because the free-energy differences between polytypes are often small, the phase observed experimentally depends sensitively on synthesis temperature, pressure, and chemical potentials. Kinetic barriers may inhibit structural transformation during cooling and allow metastable phases to persist. This interplay between thermodynamic driving forces and kinetic constraints is particularly evident in telluride systems, where subtle stacking rearrangements separate centrosymmetric and noncentrosymmetric structures~\cite{Soluyanov_2015}.

Defects provide a second axis of sensitivity linking synthesis to physical properties. Native point defects such as chalcogen vacancies, metal vacancies, antisites, and stacking faults are intrinsic products of the growth environment. Their equilibrium concentration follows
\begin{equation}
	n \propto \exp\!\left(-\frac{E_\mathrm{f}(\mu_\mathrm{M},\mu_\mathrm{X})}{k_\mathrm{B}T}\right),
\end{equation}
where the formation energy $E_\mathrm{f}$ depends explicitly on the chemical potentials of the constituent elements~\cite{Zhang_1991,VanDeWalle_2004}. First-principles calculations and spectroscopy studies demonstrate that sulfur and selenium vacancies in MoS$_2$ and related compounds introduce localized states within the band gap and strongly influence carrier density and recombination dynamics~\cite{Komsa_2015,Hong_2015}. In metallic systems, disorder modifies scattering rates and residual resistivity, while in CDW compounds it alters coherence length and transition temperature~\cite{Wilson_1975}. In topological semimetals, small stoichiometric deviations shift the Fermi level relative to band crossings and modify transport signatures~\cite{Ali_2014}.

The renewed interest in layered materials following the isolation of graphene~\cite{Novoselov_2004} accelerated the development of scalable synthesis methods for TMDs. Mechanical exfoliation enabled exploration of intrinsic electronic properties, whereas chemical vapor deposition and molecular beam epitaxy extended control to wafer-scale thin films~\cite{Lee_2012,Ugeda_2016}. Grain-boundary formation and domain coalescence in large-area films introduce extended defect networks that influence transport and optical response~\cite{Huang_2011,Manzeli_2017}. Bulk single crystals grown by chemical vapor transport, flux growth, and physical vapor transport remain essential for high-precision thermodynamic and spectroscopic measurements.

Despite substantial progress, variability between samples grown under nominally similar conditions remains widespread. Differences in temperature gradients, precursor ratios, transport-agent concentration, or cooling rates can produce measurable variations in stoichiometry, vacancy concentration, and stacking order. Reported differences in CDW transition temperatures, superconducting $T_c$, residual resistivity ratios, or optical linewidths often reflect differences in growth history rather than intrinsic discrepancies in electronic structure. These observations indicate that synthesis conditions define the thermodynamic and kinetic boundary conditions under which the electronic Hamiltonian is realized.

Although numerous reviews have surveyed TMD physics and applications~\cite{Manzeli_2017,Chen_2022}, crystal growth is often presented as a collection of experimental techniques rather than as a unified thermodynamic and kinetic problem. All synthesis routes operate within a common landscape defined by chemical potentials, supersaturation, and mass transport. Thermodynamics determines the stability window for phases and defect configurations, while kinetics governs the pathway through which the system traverses this landscape during finite-time growth. Supersaturation controls nucleation density and microstructure; diffusion and interface attachment rates regulate growth velocity; and cooling history determines metastable phase retention.

The conceptual framework developed in this Review, illustrated schematically in Fig.~\ref{fig:thermo}, integrates these thermodynamic and kinetic considerations into a unified description of TMD crystal growth. By placing bulk and thin-film techniques within a common parameter space, we clarify how synthesis conditions select polytype, stoichiometry, and microstructure. We then connect these structural variables to emergent electronic phases and emphasize that growth establishes the boundary conditions for the effective Hamiltonian realized in experiment. Through this perspective, crystal growth becomes a central variable in the reproducible and controlled realization of layered quantum materials. Although this Review focuses on TMDs, the thermodynamic and kinetic framework applies broadly to layered quantum materials. The framework introduced in Fig.~\ref{fig:thermo} will be used throughout this Review to connect synthesis conditions, thermodynamic stability, kinetic pathways, and emergent electronic phases.


\section{Thermodynamic Landscape of Phase Stability}

\begin{figure*}
	\centering
	\includegraphics[width=0.90\textwidth]{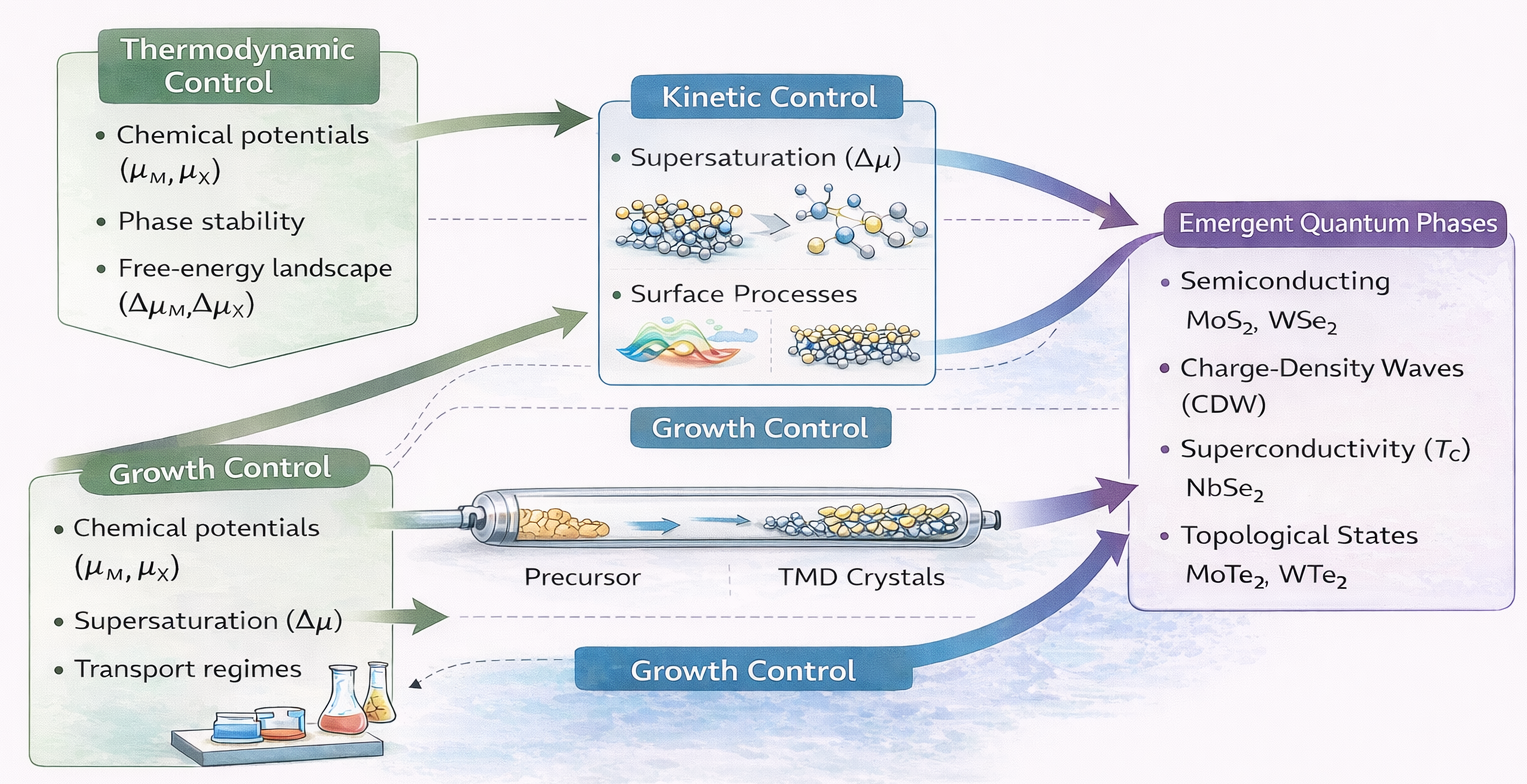}
	\caption{
		\textbf{Conceptual framework linking thermodynamic and kinetic control to emergent quantum phases in transition metal dichalcogenides.}
		Crystal growth establishes chemical potentials $(\mu_M,\mu_X)$, supersaturation $(\Delta\mu)$, and transport regimes that define the thermodynamic stability window and the kinetic pathways accessible during synthesis. Thermodynamic control governs phase stability and defect energetics through the free-energy landscape, whereas kinetic processes regulate nucleation, surface processes, and the retention of metastable phases. A representative chemical vapor transport (CVT) configuration is illustrated schematically, with precursor material located at the hot end and crystal growth occurring at the cold end of a sealed quartz ampoule. Variations in these growth parameters determine stoichiometry, defect density, and structural phase, which in turn influence the realization of semiconducting behavior (e.g., MoS$_2$, WSe$_2$), charge-density-wave order and superconductivity (e.g., NbSe$_2$), and symmetry-dependent topological states (e.g., MoTe$_2$, WTe$_2$).
	}
	\label{fig:thermo}
\end{figure*}

The equilibrium phase realized during synthesis is governed by constraints in chemical potential space. For a transition metal dichalcogenide with composition MX$_2$, thermodynamic stability can be expressed in terms of the elemental chemical potentials $\mu_M$ and $\mu_X$~\cite{Zhang_1991,VanDeWalle_2004}. Stability requires that the compound remain in equilibrium with its atomic reservoirs without precipitating elemental phases or competing secondary compounds. These conditions impose the inequalities
\begin{equation}
	\mu_M \leq \mu_M^{0}, \qquad 
	\mu_X \leq \mu_X^{0},
\end{equation}
together with the formation constraint
\begin{equation}
	\mu_M + 2\mu_X = \Delta H_f(\mathrm{MX}_2),
\end{equation}
where $\Delta H_f$ is the enthalpy of formation relative to the elemental reference states. The resulting bounded region in $(\mu_M,\mu_X)$ space defines the thermodynamic stability window of the compound. As illustrated schematically in Fig.~\ref{fig:thermo}, crystal growth selects specific points or trajectories within this stability landscape.

In practice, synthesis conditions determine the position within this stability window. Growth under chalcogen-rich conditions corresponds to $\mu_X \rightarrow \mu_X^{0}$, whereas chalcogen-poor conditions shift $\mu_X$ toward lower values with compensating changes in $\mu_M$. This distinction is particularly important for TMDs, where small deviations from stoichiometry can strongly influence defect formation and even phase selection~\cite{Manzeli_2017,Komsa_2015}. Experimentally, $\mu_\mathrm{M}$ and $\mu_\mathrm{X}$ are not directly measurable but are implicitly determined by precursor chemistry, temperature, and partial pressure.

Many TMD compositions exhibit polymorphism, with 2H, 1T, 1T$'$, and T$_\mathrm{d}$ structures separated by small free-energy differences~\cite{Chen_2022,Qian_2014}. Because these differences are often on the order of tens of meV per formula unit, modest variations in temperature, pressure, or chemical potentials can alter relative phase stability. In telluride systems such as MoTe$_2$ and WTe$_2$, the competition between 1T$'$ and T$_\mathrm{d}$ structures reflects subtle lattice distortions coupled to electronic topology~\cite{Qian_2014,Ali_2014}. In sulfides and selenides, coordination changes between trigonal prismatic and octahedral geometries also depend sensitively on thermodynamic boundary conditions~\cite{Wilson_1975,Chen_2022}.

Although equilibrium free energies determine the lowest-energy structure at fixed thermodynamic parameters, the experimentally realized phase depends on the chemical potential pathway followed during growth. Slow cooling and efficient diffusion favor transformation toward the thermodynamic ground state. Conversely, when structural rearrangements require large activation barriers, metastable polymorphs may persist. The thermodynamic landscape therefore defines the set of accessible phases, but not necessarily the one selected during synthesis.

The same chemical potential framework governs native defect formation. The formation energy of a point defect $D$ in charge state $q$ is given by~\cite{Zhang_1991,VanDeWalle_2004,Komsa_2015}
\begin{align}
	E_f(D,q) &= E_{\mathrm{tot}}(D,q) - E_{\mathrm{tot}}(\mathrm{bulk}) 
	- \sum_i n_i \mu_i  \nonumber \\
	&\quad + q(E_F + E_v) + E_{\mathrm{corr}},
\end{align}
where $n_i$ counts atoms exchanged with reservoirs of chemical potential $\mu_i$, $E_F$ is the Fermi level referenced to the valence-band maximum $E_v$, and $E_{\mathrm{corr}}$ accounts for electrostatic and finite-size corrections. The explicit dependence on $\mu_M$ and $\mu_X$ demonstrates that defect energetics are directly controlled by growth conditions.

First-principles density functional theory calculations show that chalcogen vacancies generally possess lower formation energies under chalcogen-poor conditions, leading to enhanced vacancy concentrations~\cite{Komsa_2015,Lin_2016}. Under chalcogen-rich conditions, metal vacancies or antisite defects may become competitive. Because equilibrium defect concentrations scale exponentially with formation energy, even small shifts in chemical potentials during synthesis can produce orders-of-magnitude changes in defect density. The dependence on the Fermi level also implies that charged defects can exhibit charge-transition levels within the band gap, which in turn influence doping behavior and carrier compensation in semiconducting TMDs.

These thermodynamic constraints propagate directly to electronic properties. In semiconducting TMDs, sulfur or selenium vacancies introduce donor-like states that modify carrier density and recombination dynamics~\cite{Manzeli_2017,Lin_2016}. In metallic compounds, disorder alters scattering rates and residual resistivity. In systems exhibiting charge-density-wave order or superconductivity, defect density influences coherence length and transition temperature~\cite{Wilson_1975,DiSalvo_1976}. In topological tellurides, stoichiometric deviations shift the Fermi level relative to band crossings and modify magnetotransport signatures~\cite{Ali_2014,Qian_2014}. Consequently, even modest variations in chemical potentials during synthesis can translate into large changes in defect density and measurable differences in macroscopic electronic behavior.

Thermodynamic stability therefore defines the set of allowed equilibrium configurations in phase and defect space. However, the defect population and structural phase realized during crystal growth also depend on kinetic accessibility, diffusion length scales, and supersaturation. Nonequilibrium cooling can freeze defect concentrations above equilibrium values, while limited mass transport may prevent the system from reaching the lowest free-energy configuration. A complete description of crystal growth in TMDs therefore requires integrating chemical potential constraints with kinetic pathways, which we discuss in the following section.


\section{Kinetic Pathways and Supersaturation Control}

While thermodynamics defines the region of chemical potential space in which a phase is stable, the realization of that phase during synthesis is governed by kinetics. Crystal growth requires the establishment of a supersaturated state that provides the driving force for nucleation and subsequent interface propagation. The magnitude of this driving force strongly influences crystal size, morphology, and defect incorporation.

\subsection{Supersaturation and Chemical Driving Force}

Supersaturation may be defined in terms of concentration, partial pressure, or activity, depending on the growth environment. For a generic growth process, the chemical potential difference between growth species in the reservoir and in the crystalline phase provides the thermodynamic driving force
\begin{equation}
	\Delta \mu = \mu_{\mathrm{res}} - \mu_{\mathrm{eq}},
\end{equation}
where $\mu_{\mathrm{eq}}$ corresponds to equilibrium with the solid phase. For dilute solutions or vapor phases, this expression reduces to
\begin{equation}
	\Delta \mu = k_{\mathrm{B}} T \ln S,
\end{equation}
where $S$ is the supersaturation ratio~\cite{Kashchiev_2000}. When $S=1$, the system is in equilibrium and no net growth occurs, whereas for $S>1$ nucleation and growth become thermodynamically favorable.

The magnitude of $\Delta \mu$ governs both nucleation density and growth velocity. Low supersaturation favors fewer nuclei and the formation of larger crystals, whereas high supersaturation promotes rapid nucleation and fine-grained microstructures. This behavior is common to vapor-phase, solution-based, and melt-growth methods, although the physical origin of supersaturation differs among them.

\subsection{Classical Nucleation Barrier}

\begin{figure}[h]
	\centering
	\includegraphics[width=0.9\linewidth]{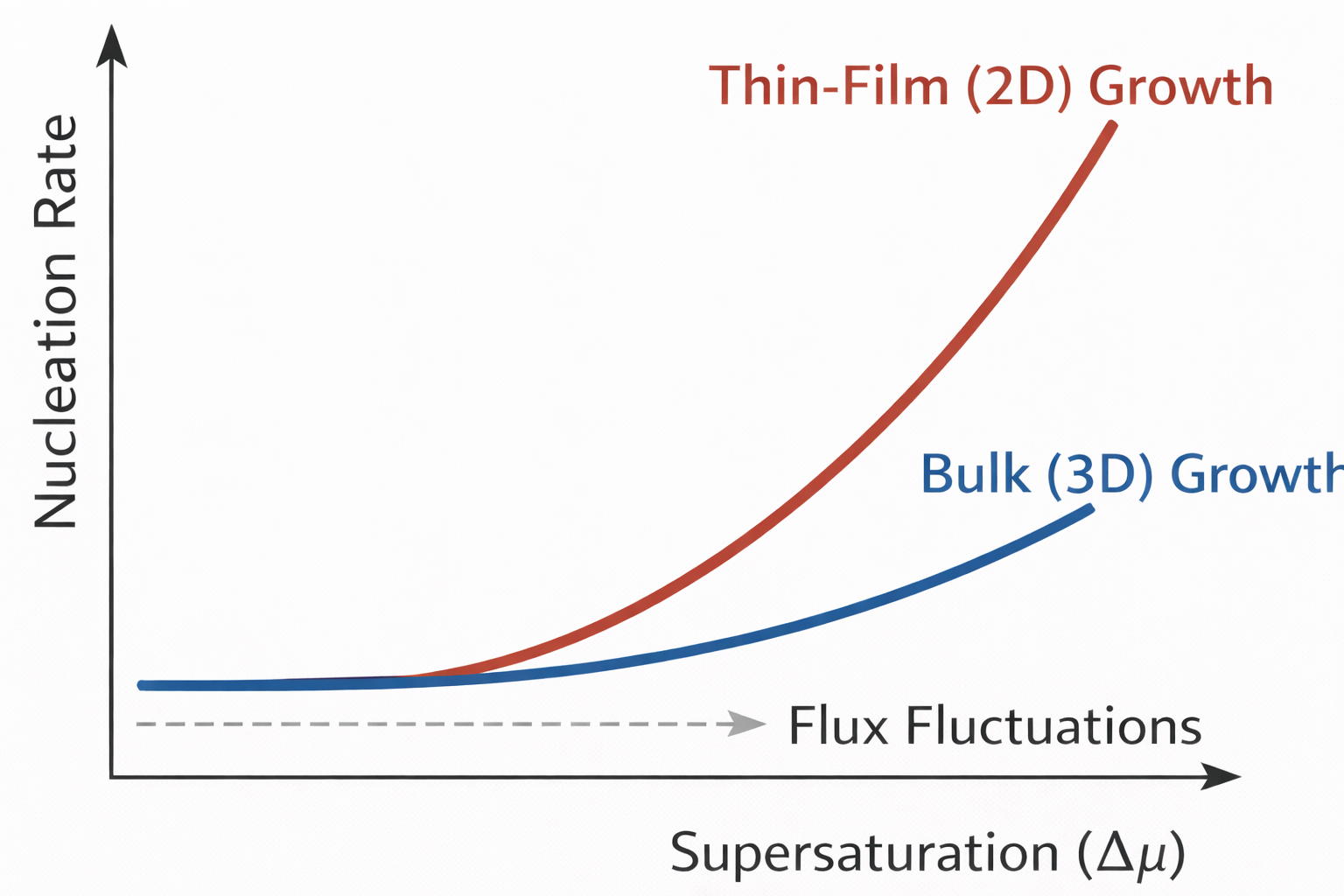}
	\caption{
		Schematic comparison of nucleation rate as a function of supersaturation 
		($\Delta \mu$) for bulk three-dimensional (3D) crystal growth and 
		two-dimensional (2D) thin-film growth. In classical nucleation theory the nucleation barrier scales as 
		$\Delta G^{*} \propto (\Delta \mu)^{-2}$ for three-dimensional nuclei, 
		so the increase in nucleation rate with supersaturation is relatively 
		gradual in bulk systems. For two-dimensional nucleation on surfaces the barrier scales approximately as $\Delta G^{*} \propto (\Delta \mu)^{-1}$, leading to a much stronger sensitivity of nucleation rate to supersaturation. Consequently, small fluctuations in precursor flux or chemical potential during thin-film growth can generate large variations in nucleation density, often producing polycrystalline films with grain boundaries.
	}
	\label{fig:nucleation_supersaturation}
\end{figure}

The formation of a new crystalline nucleus requires overcoming an energetic barrier associated with the creation of an interface between the nucleus and its surrounding environment. The nucleation rate depends exponentially on the nucleation barrier,
\begin{equation}
	J \propto \exp\!\left(-\frac{\Delta G^\ast}{k_B T}\right),
\end{equation}
so reductions in $\Delta G^\ast$ caused by increased supersaturation can lead to rapid increases in nucleation density. For a spherical nucleus of radius $r$ and interfacial energy $\gamma$, the free-energy change is
\begin{equation}
	\Delta G(r) = - \frac{4}{3}\pi r^3 \frac{\Delta \mu}{v_m}
	+ 4\pi r^2 \gamma,
\end{equation}
where $v_m$ is the molecular volume~\cite{Kashchiev_2000}. Competition between the volume free-energy gain and the surface-energy cost yields a critical radius
\begin{equation}
	r^\ast = \frac{2\gamma v_m}{\Delta \mu},
\end{equation}
and a corresponding nucleation barrier
\begin{equation}
	\Delta G^\ast =
	\frac{16\pi \gamma^3 v_m^2}{3 (\Delta \mu)^2}.
\end{equation}

Because $\Delta G^\ast \propto (\Delta \mu)^{-2}$, even modest increases in supersaturation substantially reduce the nucleation barrier. This inverse-square dependence explains the strong sensitivity of nucleation density to temperature gradients, precursor flux, and local chemical potential variations during TMD growth. Heterogeneous nucleation on substrates, ampoule walls, or seed crystals reduces the geometric prefactor of the barrier, although the scaling with $\Delta \mu$ remains qualitatively unchanged.

The sensitivity of nucleation kinetics to supersaturation differs significantly between bulk and thin-film growth. This contrast is illustrated schematically in Fig.~\ref{fig:nucleation_supersaturation}, which compares the dependence of nucleation rate on supersaturation for three-dimensional and two-dimensional growth. Because the nucleation barrier depends strongly on supersaturation, even small fluctuations in precursor flux or local chemical potential during growth can lead to large variations in nucleation density.

\subsection{Mass Transport and Interface Kinetics}

Following nucleation, sustained crystal growth requires transport of growth species to the interface and their incorporation into the lattice. The overall growth rate is determined by the slower of two processes: mass transport through the surrounding medium or attachment kinetics at the interface.

In diffusion-limited growth, the flux to the interface may be approximated as
\begin{equation}
	J_{\mathrm{diff}} \approx D \frac{C - C_{\mathrm{eq}}}{\delta},
\end{equation}
where $D$ is the diffusivity, $C$ is the bulk concentration, $C_{\mathrm{eq}}$ is the equilibrium concentration, and $\delta$ is an effective diffusion boundary-layer thickness~\cite{Bird_2002}. Enhanced convection or larger transport geometries reduce $\delta$ and increase the mass flux.

In interface-limited growth, incorporation at the crystal surface becomes rate controlling. A commonly used expression for the interface velocity is
\begin{equation}
	v = v_0 
	\left[1 - \exp\!\left(-\frac{\Delta \mu}{k_{\mathrm{B}}T}\right)\right],
\end{equation}
where $v_0$ contains kinetic prefactors associated with surface diffusion and attachment frequency~\cite{Burton_1951}. In the limit $\Delta \mu \ll k_{\mathrm{B}}T$, the growth velocity scales approximately linearly with supersaturation.

The competition between transport and interface kinetics can be characterized by a dimensionless parameter analogous to a Damk\"ohler number,
\begin{equation}
	\mathrm{Da} = \frac{k_s L}{D},
\end{equation}
where $k_s$ is an effective surface reaction rate constant, $L$ is a characteristic transport length, and $D$ is the diffusivity. For $\mathrm{Da} \gg 1$, growth is diffusion limited, whereas for $\mathrm{Da} \ll 1$, growth is interface limited. This dimensionless parameter provides a convenient way to compare otherwise disparate growth techniques within a unified kinetic framework.

\subsection{Kinetic Trapping and Metastability}

When cooling rates are finite and diffusion lengths are limited, the system may not reach the equilibrium configuration predicted by thermodynamics. If structural transformations between polytypes require collective atomic rearrangements with substantial activation barriers, a high-temperature phase can persist upon cooling. Rapid quenching can similarly freeze defect populations at concentrations exceeding their equilibrium values at room temperature.

These kinetic constraints are particularly relevant in TMDs with small free-energy differences between competing polytypes or strong coupling between lattice distortion and electronic structure. In such systems, synthesis parameters including temperature gradient, precursor flux, total pressure, and cooling rate determine whether the material approaches equilibrium or remains trapped in a metastable configuration.

Supersaturation, transport regime, and kinetic barriers therefore act together with thermodynamic stability constraints to define the accessible structural and defect landscape during crystal growth. In the next section we apply this thermodynamic and kinetic framework to bulk crystal growth methods and identify the regimes in which specific techniques operate.

An important manifestation of kinetic trapping occurs during the cooling stage of crystal growth. If cooling proceeds slowly, atomic diffusion and structural rearrangements may allow the system to transform toward the thermodynamically stable phase predicted by equilibrium free-energy considerations. In contrast, rapid cooling or quenching can suppress long-range atomic rearrangements and effectively freeze the high-temperature structure into a metastable configuration. This mechanism is particularly relevant for transition metal dichalcogenides exhibiting small energy differences between competing polytypes.

\begin{figure}[h]
	\centering
	\includegraphics[width=0.9\linewidth]{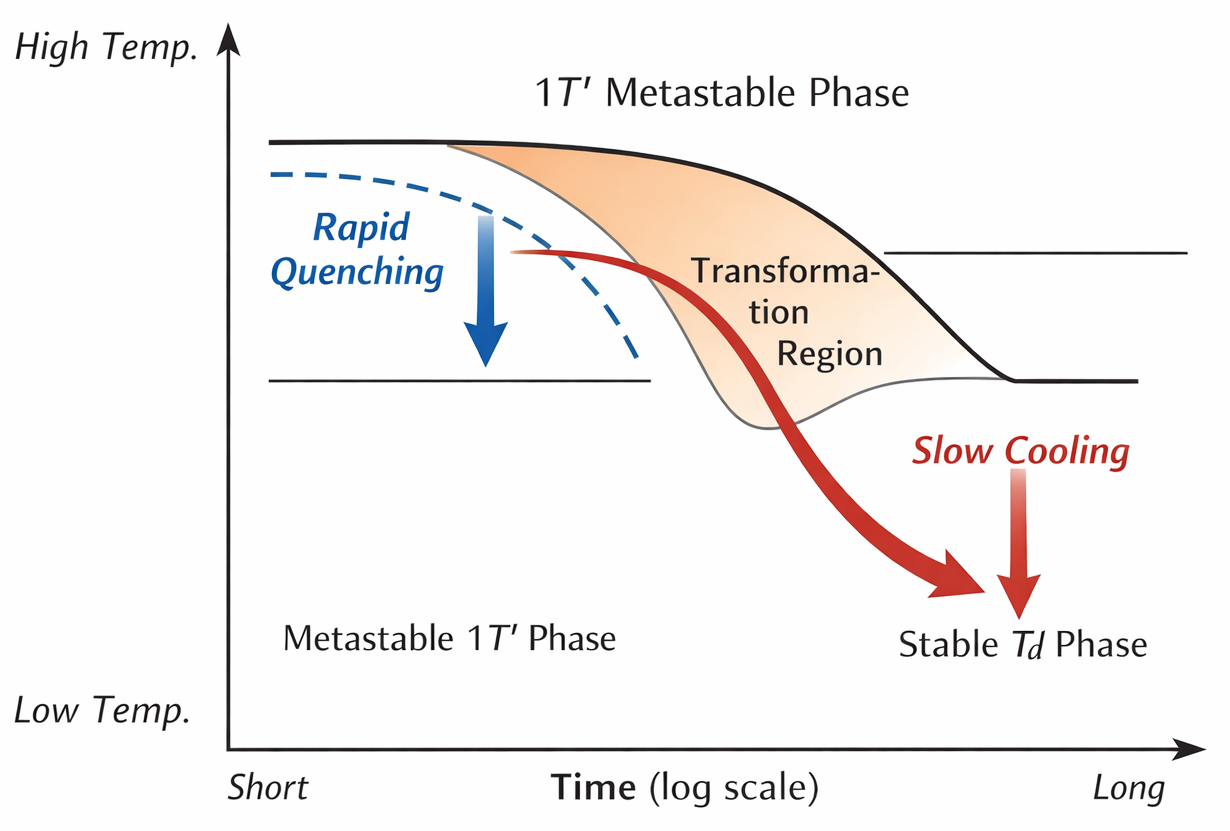}
	\caption{
		Schematic time–temperature–transformation (TTT) diagram illustrating kinetic trapping 
		of metastable phases in transition metal dichalcogenides, exemplified by MoTe$_2$. 
		The curve represents the characteristic time required for transformation between 
		competing polytypes (e.g., $1T'$ and $T_d$). During slow cooling, the system crosses 
		the transformation boundary and evolves toward the thermodynamically stable phase. 
		In contrast, rapid quenching bypasses the transformation region, preserving the 
		high-temperature structure as a metastable phase at low temperature. Such kinetic 
		trapping can determine the symmetry and electronic topology realized in experiment.
	}
	\label{fig:TTT}
\end{figure}

A representative example is MoTe$_2$~\cite{Qian_2014, Soluyanov_2015}, where the centrosymmetric $1T'$ phase and the noncentrosymmetric $T_d$ phase differ in energy by only tens of meV per formula unit (Fig.~\ref{fig:TTT}). During growth or post-growth cooling, transformation between these structures requires coordinated lattice distortions and atomic displacements. If the cooling rate exceeds the characteristic timescale for structural transformation, the high-temperature phase may persist to low temperature. Such kinetic trapping can stabilize structures that would otherwise be thermodynamically unfavorable and thereby alter the symmetry and electronic topology of the resulting material.

The competition between thermodynamic driving forces and kinetic barriers during cooling can be illustrated using a time–temperature–transformation (TTT) diagram, which identifies the temperature and time scales over which structural transformations become kinetically accessible.


\section{Bulk Crystal Growth in the Thermodynamic--Kinetic Map}

\begin{figure*}
	\centering
	\includegraphics[width=0.90\textwidth]{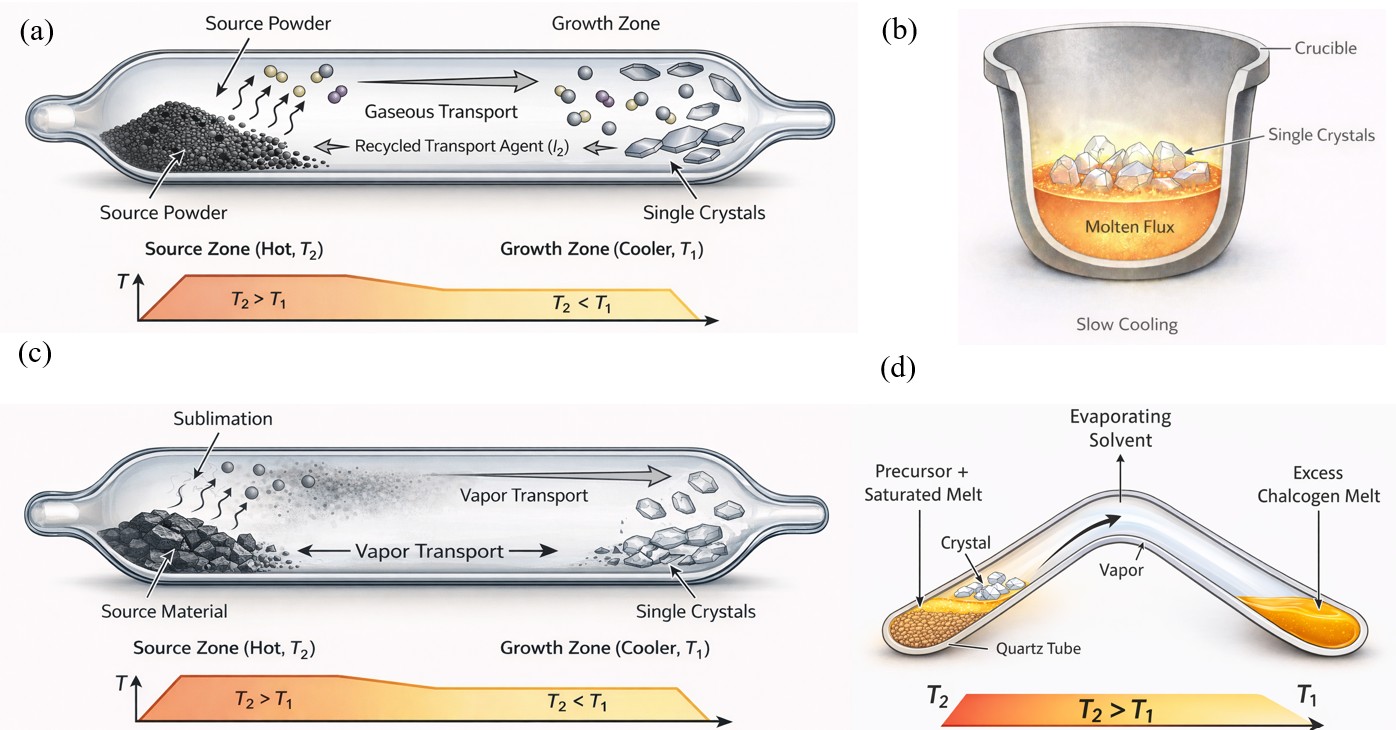}
	\caption{
		\textbf{Representative bulk crystal growth techniques for transition metal dichalcogenides.}
		(a) Chemical vapor transport (CVT). A controlled temperature gradient establishes distinct hot ($T_\mathrm{H}$) and cold ($T_\mathrm{L}$) zones within a sealed ampoule. In the hot zone, reversible reactions between the precursor and a transport agent form volatile species that migrate along the gradient. Decomposition in the colder region generates supersaturation and drives nucleation and single-crystal growth.
		(b) Flux growth. Precursor materials dissolve in a molten solvent at elevated temperature, forming a homogeneous solution. Controlled cooling or gradual supersaturation of the melt induces crystallization under near-equilibrium conditions.
		(c) Physical vapor transport (PVT). Material sublimates in the high-temperature region ($T_2$) and recondenses in a lower-temperature zone ($T_1$) under a thermal gradient ($T_2 > T_1$). Mass transport proceeds through intrinsic vapor species without the use of a transport agent.
		(d) Chalcogen-melt solvent-evaporation method. A metal-saturated chalcogen melt is maintained at high temperature ($T_2$). Excess chalcogen vapor migrates toward a cooler region ($T_1$), where it condenses, progressively shifting the melt composition toward metal supersaturation and promoting crystal growth in the hot zone.
	}
	\label{fig:growth}
\end{figure*}

Bulk single crystals remain essential for transport, thermodynamic, and spectroscopic investigations of transition metal dichalcogenides. Although chemical vapor transport, flux growth, physical vapor transport, and solvent-assisted crystallization differ operationally, each method occupies a distinct region of chemical potential space and kinetic parameter space. Figure~\ref{fig:growth} summarizes representative configurations of these approaches. In this section we analyze these techniques within the unified thermodynamic and kinetic framework developed in the previous sections.

\subsection{Chemical Vapor Transport}

Chemical vapor transport (CVT) is widely used for growing high-quality TMD single crystals~\cite{Binnewies_2017}. In CVT, a transport agent such as iodine or bromine forms volatile intermediates with the transition metal, enabling material transport along a temperature gradient. The process can be described schematically by coupled equilibrium reactions

\begin{equation}
	\mathrm{MX_2 (s)} + \mathrm{X_2 (g)} \rightleftharpoons \mathrm{MX_{2+y} (g)},
\end{equation}

followed by decomposition in the colder region of the ampoule.

Transport occurs through volatile metal–halide complexes whose equilibrium partial pressures determine the direction and magnitude of material transport along the temperature gradient. Within chemical potential space, CVT modifies the effective $\mu_M$ and $\mu_X$ through the partial pressures of volatile complexes. The imposed temperature gradient generates a spatial variation in supersaturation that is typically largest near the growth zone. Because transport occurs in the vapor phase, growth may become diffusion limited when the flux of transport species controls mass delivery. The resulting kinetic regime depends sensitively on the temperature difference, transport-agent concentration, and ampoule geometry.

CVT commonly operates at moderate supersaturation, which promotes large crystal size and good structural order. However, halogen species can influence defect equilibria and introduce impurity incorporation. In addition, spatial variations in $\mu_X$ along the temperature gradient may alter vacancy formation energies and local stoichiometry.

An additional consideration in CVT growth is the possible incorporation of transport-agent species into the crystal lattice. Halogens such as iodine or bromine are used to form volatile metal--halide complexes that enable vapor-phase transport. Although the transport agent ideally remains in the gas phase, trace incorporation of halogen atoms can occur during crystal growth. The electronic consequences of such impurities depend on their lattice site and concentration. In some cases, halogen substitution or interstitial incorporation may introduce donor-like states that modify the carrier concentration and act as weak dopants. More commonly, however, these impurities behave as scattering centers that increase disorder and reduce carrier mobility.

Experimental studies on metallic TMDs have shown that residual halogen impurities can reduce the residual resistivity ratio (RRR), reflecting enhanced impurity scattering at low temperature. Because collective electronic states such as superconductivity or charge-density-wave order are sensitive to disorder, even small concentrations of halogen impurities can influence transition temperatures and transport behavior~\cite{Binnewies_2017,Wilson_1975,Canfield_1992}. For experiments requiring ultra-high-purity crystals, transport-agent-free methods such as physical vapor transport (PVT) may therefore be preferable, despite their typically lower growth rates and smaller crystal sizes.

This trade-off between growth efficiency and chemical purity highlights the importance of selecting growth techniques that are appropriate for the targeted electronic properties.

\subsection{Flux Growth}

Flux growth employs a molten solvent to dissolve precursor materials and induce crystallization upon cooling~\cite{Canfield_1992}. In this approach, chemical potentials are determined by solution thermodynamics rather than vapor equilibria. Supersaturation develops gradually as the temperature decreases below the liquidus.

Because diffusion in liquids is relatively efficient and convection may occur, flux growth often approaches the interface-limited regime at low supersaturation. This suppresses nucleation density and favors the formation of large, well-faceted crystals. Growth closer to thermodynamic equilibrium can also reduce nonequilibrium defect incorporation.

The accessible region of chemical potential space is constrained by the solvent composition and the corresponding phase diagram. Flux components may incorporate into the lattice or form secondary phases if solubility limits are exceeded. Flux growth therefore balances improved crystallinity against restricted compositional flexibility.

\subsection{Physical Vapor Transport}

Physical vapor transport (PVT) relies on sublimation of the source material and recondensation under a temperature gradient without chemical transport agents~\cite{Rudolph_2003}. Vapor species consist primarily of intrinsic fragments of the compound, and chemical potentials are set by intrinsic vapor pressures.

PVT typically operates at elevated temperatures where vapor pressures are sufficient to sustain transport. Supersaturation is controlled by the temperature difference between source and growth zones. Because transport does not rely on reactive intermediates, stoichiometry can often be preserved more directly than in halogen-assisted CVT. However, enhanced chalcogen volatility at high temperature can shift $\mu_X$ toward chalcogen-poor conditions and thereby increase vacancy concentrations.

The kinetic regime depends on pressure and temperature. Under low-pressure conditions diffusion may be efficient and growth becomes interface limited, whereas at higher pressures diffusion resistance can dominate. PVT therefore spans a broad region of the thermodynamic and kinetic map depending on operating parameters.

\begin{table*}[htbp]
	\centering
	\caption{Comparison of major bulk single-crystal growth techniques for transition metal dichalcogenides}
	\label{tab:bulk_comparison}
	
	\begin{tblr}{
			colspec = {p{2.5cm} p{5.0cm} p{5.0cm} p{4.5cm}},
			row{1} = {font=\bfseries},
			hlines,
			cells = {t},               
			abovesep = 4pt,
			belowsep = 4pt,
			stretch = 1.5,             
		}
		Technique & Key advantages & Major challenges & Representative materials \\
		\hline
		CVT & 
		Large, high-quality single crystals; excellent crystallinity; access to metastable polytypes; widely applicable to layered TMDs & 
		Transport-agent incorporation; sensitivity to temperature gradients; limited control over stoichiometry; long growth times & 
		MoS$_2$, WS$_2$, MoSe$_2$, WSe$_2$, TiSe$_2$, NbSe$_2$~\cite{Lv_2015,Binnewies_2017,Wilson_1975} \\
		Flux growth & 
		Low growth temperatures; growth of air-sensitive compounds; large crystal thickness; reduced vapor-phase defects & 
		Flux inclusion and contamination; difficult flux removal; limited material solubility; composition deviations & 
		NiTe$_2$, CoSb$_2$, NbS$_2$, TaS$_2$~\cite{Je_2023,Laz_2012,Nagao_2021,Zhu_2009,Colell_1994,Canfield_2016} \\
		PVT & 
		High-purity crystals; no transport agents required; simple experimental setup; minimal chemical contamination & 
		High growth temperatures; small crystal size; low yield; sensitivity to vapor pressure balance & 
		MoTe$_2$, WTe$_2$, HfTe$_2$, ZrTe$_2$~\cite{Zhang_2020,Colell_1994} \\
		Solvent evaporation & 
		Low-temperature synthesis; cost-effective and scalable; easy compositional tuning; suitable for exploratory growth & 
		Limited crystal size; variable crystallinity; solvent residues; phase purity challenges & 
		PtTe$_2$, PdTe$_2$, TiTe$_2$~\cite{Chareev_2020,Zhang_2020} \\
		\hline
	\end{tblr}
\end{table*}

\subsection{Solvent-Assisted and Melt Growth}

In systems that melt congruently or form stable liquid solutions, growth from a self-flux or near-stoichiometric melt is possible~\cite{Rudolph_2003}. Chemical potentials are determined by melt composition and temperature, and supersaturation develops upon cooling below the liquidus.

Melt-based growth can provide relatively homogeneous chemical potentials and reduced spatial gradients compared with vapor-phase methods. However, many TMDs melt incongruently or decompose before forming stable liquids, which limits the applicability of melt-growth techniques. Phase-diagram constraints must therefore be carefully evaluated to avoid the formation of secondary phases. The choice of growth technique therefore determines not only crystal size and morphology but also the accessible chemical potential window and defect landscape, which ultimately influence the electronic phases realized in experiment.

\subsection{Mapping Bulk Methods onto the Thermodynamic--Kinetic Framework}

Table~\ref{tab:bulk_comparison} summarizes the principal strengths and limitations of each method. Interpreted within the thermodynamic and kinetic framework, these techniques correspond to distinct combinations of supersaturation, transport regime, and chemical potential control.

Flux growth generally occupies the low-supersaturation, near-equilibrium region and favors large crystals with reduced nonequilibrium defect populations. CVT spans moderate supersaturation with vapor-phase diffusion control and adjustable chemical potentials through transport-agent chemistry. PVT accesses high-temperature regimes where intrinsic vapor pressures define both $\mu_X$ and the kinetic transport balance. Solvent-assisted methods operate in compositionally constrained regions determined by melt thermodynamics.

Bulk growth techniques can therefore be understood as controlled trajectories through a common thermodynamic and kinetic landscape rather than unrelated experimental procedures. In the next section we extend this analysis to thin-film growth, where reduced dimensionality and substrate interactions introduce additional constraints on nucleation, strain, and phase stability.


\section{Thin-Film Growth and Reduced Dimensionality Effects}

Thin-film synthesis of transition metal dichalcogenides introduces constraints that are absent in bulk growth. Nucleation occurs on a substrate, interfacial and strain energies compete with the bulk free energy, and growth conditions are often far from thermodynamic equilibrium. Substrate interactions, strain, and two-dimensional nucleation kinetics therefore play central roles in determining phase stability and microstructure.

\subsection{Two-Dimensional Nucleation}

For epitaxial or van der Waals thin-film growth, nucleation proceeds in two dimensions. The free-energy change associated with forming a circular nucleus of radius $r$ on a surface is

\begin{equation}
	\Delta G(r) = - \pi r^2 \frac{\Delta \mu}{\Omega}
	+ 2 \pi r \beta,
\end{equation}

where $\Omega$ is the area per growth unit and $\beta$ is the step-edge free energy~\cite{Venables_1984}. Reduced dimensionality modifies the balance between thermodynamic driving force and interfacial energy compared with bulk nucleation. The critical radius becomes

\begin{equation}
	r^\ast = \frac{\beta \Omega}{\Delta \mu},
\end{equation}

and the nucleation barrier scales as $(\Delta \mu)^{-1}$ rather than $(\Delta \mu)^{-2}$.

This weaker power-law dependence increases the sensitivity of nucleation density to supersaturation in two dimensions. Thin-film growth often begins at relatively high effective supersaturation, particularly during precursor exposure or flux ramping. As a result, nucleation densities can become large, producing polycrystalline films with grain boundaries. Careful control of precursor flux, substrate temperature, and background pressure is therefore required to achieve large single-domain areas.

\subsection{Substrate Interaction and Strain}

In contrast to bulk crystals, thin films experience interfacial interactions with the substrate. Even in nominal van der Waals epitaxy, residual lattice mismatch and thermal expansion differences introduce strain. The total free energy of a thin film may be expressed schematically as

\begin{equation}
	G_{\mathrm{film}} = G_{\mathrm{bulk}} + G_{\mathrm{interface}} + G_{\mathrm{strain}},
\end{equation}

where $G_{\mathrm{interface}}$ represents adhesion energy and $G_{\mathrm{strain}}$ accounts for elastic strain contributions.

For TMDs with small free-energy differences between competing polytypes, such as MoTe$_2$ or WTe$_2$, strain can stabilize one phase relative to another~\cite{Qian_2014}. Substrate-induced symmetry breaking may influence stacking order, crystallographic orientation, and electronic topology. Phase selection in thin films therefore reflects a competition between intrinsic thermodynamic stability and extrinsic interfacial constraints.

\begin{figure}[h]
	\centering
	\includegraphics[width=0.5\textwidth]{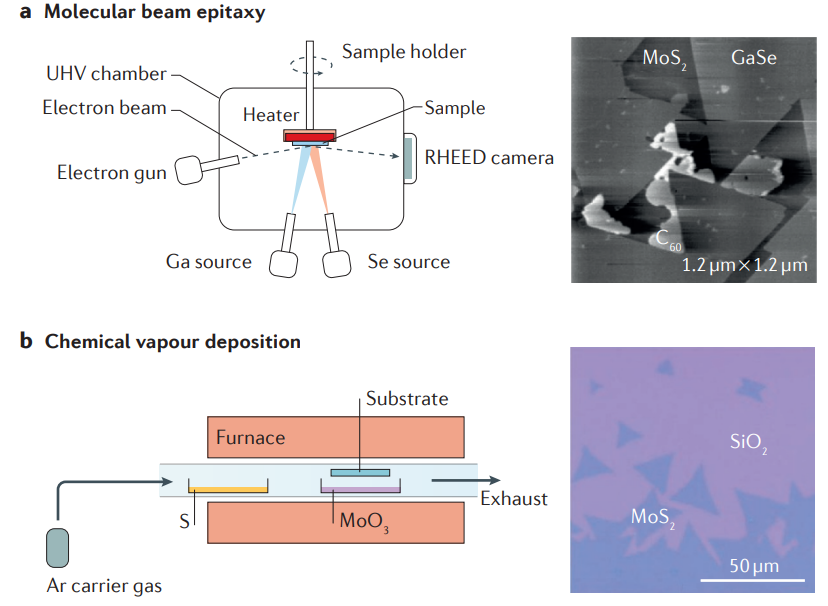}
	\caption{
		\textbf{Representative growth approaches for ultrathin transition metal dichalcogenides (TMDCs).}
		(a) In molecular beam epitaxy (MBE), high-purity elemental sources are thermally evaporated under ultrahigh vacuum, enabling independent control of metal and chalcogen fluxes during layer-by-layer growth. The right panel shows an atomic force microscopy image of monolayer GaSe epitaxially grown on MoS$_2$. 
		(b) In chemical vapor deposition (CVD), reactive precursor species are transported in the vapor phase and undergo chemical reactions at the substrate surface, leading to nucleation and lateral domain expansion. The right panel displays an optical micrograph of triangular single-crystalline MoS$_2$ domains synthesized on SiO$_2$. 
		Reproduced with permission from Ref.~\cite{Manzeli_2017}.
	}
	\label{fig:mbe}
\end{figure}

Figure~\ref{fig:mbe} illustrates representative thin-film growth environments. In contrast to the bulk methods summarized in Fig.~\ref{fig:growth}, thin-film techniques operate under dynamic flux control and substrate-mediated boundary conditions.

Beyond elastic strain, the dielectric environment provided by the substrate can substantially modify defect energetics in two-dimensional TMDs. In the monolayer limit, reduced dielectric screening enhances Coulomb interactions associated with charged defects, making defect formation energies sensitive to the surrounding electrostatic environment. When a monolayer is supported on a substrate, the effective dielectric screening increases relative to the free-standing case. This increased screening can lower the formation energies of charged defects by stabilizing their electrostatic potential~\cite{Raja_2017,Lin_2016}. The substrate therefore acts as an additional thermodynamic boundary condition that modifies defect energetics and electronic structure in two-dimensional TMD systems. This substrate dependence becomes particularly important in ultrathin films, where electrostatic screening is dominated by the surrounding dielectric environment rather than by the material itself.

First-principles calculations demonstrate that this effect depends strongly on substrate properties~\cite{Komsa_2015}. For example, monolayer MoS$_2$ supported on SiO$_2$ can exhibit lower formation energies for sulfur vacancies compared with the free-standing case because of enhanced dielectric screening and possible charge transfer at the interface. In contrast, hexagonal boron nitride (h-BN), which provides a cleaner and more weakly interacting van der Waals interface, preserves intrinsic defect energetics more closely while suppressing extrinsic disorder. As a result, TMD monolayers grown or transferred onto h-BN typically exhibit improved carrier mobility and narrower optical linewidths compared with those supported on SiO$_2$. These substrate-dependent changes in defect energetics illustrate that the dielectric environment constitutes an additional control parameter linking growth conditions, defect populations, and electronic properties in two-dimensional TMD systems.

Because defect concentrations scale exponentially with formation energy, even modest substrate-induced changes in dielectric screening can produce orders-of-magnitude variations in equilibrium defect density. Such variations can significantly influence carrier density, scattering rates, and optical response in two-dimensional TMD materials.

\begin{table*}[htbp]
	\centering
	\caption{Comparison of major vapor-phase thin-film growth techniques for transition metal dichalcogenides}
	\label{tab:thinfilm_comparison}
	
	\begin{tblr}{
			colspec = {p{2.5cm} p{5.0cm} p{5.0cm} p{4.5cm}},
			row{1} = {font=\bfseries},
			hlines,
			cells = {t},             
			abovesep = 4pt,
			belowsep = 4pt,
			stretch = 1.5,           
		}
		Technique & Key advantages & Major challenges & Representative materials \\
		\hline
		CVD & 
		Large-area growth; wafer-scale scalability; monolayer thickness control; substrate compatibility & 
		Grain boundaries; high nucleation density; point defects; strain inhomogeneity & 
		MoS$_2$, WS$_2$, MoSe$_2$, WSe$_2$~\cite{Lee_2012,VanDerZande_2013,Manzeli_2017} \\
		MBE & 
		Precise stoichiometry control; ultra-high vacuum purity; epitaxial film growth; atomically sharp interfaces & 
		Chalcogen vacancies; low growth rates; small domain sizes; limited scalability & 
		MoSe$_2$, WSe$_2$, NbSe$_2$, VSe$_2$~\cite{Xue_2024,Cheng_2020,Suresh_2018} \\
		PLD & 
		Stoichiometric transfer from target; flexible target composition; rapid materials exploration; multilayer heterostructures & 
		High defect density; surface roughness; chalcogen re-evaporation; particulate formation & 
		MoS$_2$, WS$_2$, NbSe$_2$~\cite{Singh_2020,Rai_2020} \\
		Sputtering & 
		Industrial scalability; CMOS compatibility; uniform large-area deposition & 
		Requires post-annealing; extensive post-annealing; limited phase control & 
		MoS$_2$, WS$_2$~\cite{Wong_2017,Husaain_2016} \\
		\hline
	\end{tblr}
\end{table*}

\subsection{Chemical Vapor Deposition}

Chemical vapor deposition (CVD) is widely used for large-area growth of monolayer and few-layer TMDs~\cite{Lee_2012}. Precursor decomposition generates reactive species that diffuse to the substrate and incorporate into the growing lattice. Chemical potentials are determined dynamically by precursor fluxes, carrier gas flow, and substrate temperature rather than by static equilibrium conditions.

CVD typically operates with relatively high supersaturation during the initial nucleation stage, followed by reduced effective supersaturation during lateral domain growth. This produces isolated triangular or hexagonal domains that expand until coalescence. Grain boundaries formed during domain merging strongly influence electronic transport and optical response. Their density is primarily determined by early-stage nucleation kinetics and substrate chemistry.

\subsection{Molecular Beam Epitaxy}

Molecular beam epitaxy (MBE) provides independent control of metal and chalcogen fluxes under ultrahigh vacuum conditions~\cite{Ugeda_2016}. Low flux and controlled substrate temperature can place growth near the interface-limited regime, enabling layer-by-layer deposition.

Independent tuning of $\mu_M$ and $\mu_X$ allows systematic exploration of chalcogen-rich and chalcogen-poor conditions. Insufficient chalcogen flux promotes vacancy formation, whereas excessive flux may lead to desorption-limited growth or secondary phase formation. The balance between surface diffusion, desorption, and incorporation ultimately determines film morphology, stoichiometry, and defect density.

\subsection{Grain Boundaries and Dimensional Constraints}

Table~\ref{tab:thinfilm_comparison} highlights the trade-offs among major thin-film growth techniques. High-supersaturation methods such as CVD and PLD enable rapid large-area deposition but often produce high nucleation densities and grain boundaries. Techniques operating closer to the interface-limited regime, such as MBE, offer improved stoichiometric control but at reduced growth rates and scalability.

Grain boundaries, rotational domains, and thickness nonuniformity are intrinsic consequences of two-dimensional nucleation and domain coalescence. In TMD monolayers, grain boundaries can host metallic states or localized defect complexes that modify transport and optical response~\cite{Manzeli_2017}. Their density is primarily determined during early growth stages when supersaturation and adatom mobility determine the nucleation rate.

Reduced dimensionality also modifies intrinsic defect energetics. Formation energies in monolayers can differ from bulk values because of reduced dielectric screening and modified bonding environments~\cite{Komsa_2015}. Consequently, thin films grown under nominally similar chemical potentials may exhibit defect landscapes that differ significantly from those of bulk single crystals.

\subsection{Thin Films within the Thermodynamic--Kinetic Framework}

Thin-film growth occupies a region of parameter space characterized by two-dimensional nucleation, significant interface contributions, and dynamically controlled chemical potentials. Supersaturation is often higher and more transient than in bulk growth, and nonequilibrium effects are therefore more pronounced.

Within the unified thermodynamic and kinetic framework, thin-film methods can be viewed as operating in reduced dimensionality with additional interfacial energy terms superimposed on the bulk stability landscape defined in Section II. This perspective clarifies systematic differences in phase selection, defect density, and microstructure between bulk crystals and thin films.

In the next section we combine bulk and thin-film perspectives to establish a direct connection between growth conditions, defect populations, and emergent electronic phases.


\section{Linking Growth, Defects, and Emergent Quantum Phases}

The thermodynamic and kinetic principles developed in the preceding sections establish how growth conditions determine chemical potentials, supersaturation, and defect populations. We now examine how these structural variables influence emergent electronic phases in transition metal dichalcogenides. In many TMDs, collective phenomena arise from small free-energy differences. Consequently, modest variations in stoichiometry or disorder introduced during synthesis can shift phase boundaries and modify low-temperature behavior.

\subsection{Charge-Density-Wave Order}

Several TMDs exhibit charge-density-wave (CDW) transitions driven by electron--phonon coupling and Fermi-surface instabilities~\cite{Wilson_1975,DiSalvo_1976}. The coherence of the CDW state requires lattice periodicity and long-range electronic order. Point defects and compositional disorder break translational symmetry, reduce correlation length, and broaden thermodynamic and spectroscopic signatures of the transition.

In 1T-TiSe$_2$, small stoichiometric variations modify the balance between semimetallic band overlap and excitonic instability~\cite{DiSalvo_1976}. Chalcogen deficiency or intercalation alters carrier concentration and shifts the CDW phase boundary. In 2H-NbSe$_2$, disorder influences the competition between CDW order and superconductivity~\cite{Wilson_1975}. Growth under chalcogen-poor conditions increases vacancy density, enhancing scattering and reducing CDW coherence length.

Because vacancy formation energies depend explicitly on the chalcogen chemical potential $\mu_X$, growth methods that maintain chalcogen-rich conditions tend to suppress vacancy concentrations and stabilize long-range CDW order. In contrast, chalcogen-deficient or high-temperature growth conditions increase disorder and can modify or suppress collective ordering.

\begin{figure*}
	\centering
	\includegraphics[width=0.9\textwidth]{SC.png}
	\caption{
		\textbf{Disorder-enhanced superconductivity in monolayer NbSe$_2$.}
		(a) Scanning tunneling microscopy (STM) topography of a NbSe$_2$ monolayer with randomly distributed Si adatoms (14\,nm $\times$ 14\,nm; sample bias 1.0\,V; tunneling current 100\,pA). Bright protrusions correspond to individual Si adatoms. 
		(b) Differential conductance ($dI/dV$) spectra acquired near the Fermi level at base temperature (set point: 1\,mV, 100\,pA) for increasing Si adatom density, showing systematic modification of the superconducting gap. 
		(c) Extracted superconducting gap magnitude as a function of Si adatom concentration; the dashed line serves as a guide to the eye. 
		(d) Atomically resolved STM image of NbSe$_{2-x}$S$_x$ (x = 0.31) monolayer (10\,nm $\times$ 10\,nm; 100\,mV; 100\,pA), where sulfur substitution appears as darker atomic sites. 
		(e) $dI/dV$ spectra at base temperature for varying sulfur content in NbSe$_{2-x}$S$_x$, illustrating the evolution of the superconducting gap with alloy disorder. 
		(f) Superconducting transition temperature as a function of sulfur concentration $x$. Open squares denote the gap-closing temperature, while red circles indicate the disappearance of coherence peaks; dashed lines are guides to the eye. 
		Reproduced with permission from Ref.~\cite{Zhao_2019}.
	}
	\label{fig:sc}
\end{figure*}

\subsection{Superconductivity and Disorder}

Superconductivity in metallic TMDs is sensitive to defect density. In 2H-NbSe$_2$ and related compounds, the superconducting transition temperature $T_c$ depends on the electron--phonon coupling strength and the density of states at the Fermi level. Defects introduced during growth modify scattering rates and influence pairing interactions.

Figure~\ref{fig:sc} illustrates the controlled tuning of superconductivity in monolayer NbSe$_2$ through disorder~\cite{Zhao_2019}. Introducing adatoms or alloy disorder systematically modifies the superconducting gap and transition temperature. In systems where superconductivity coexists or competes with CDW order, defect density shifts the balance between these phases. Moderate disorder can weaken CDW order, whereas strong scattering may suppress $T_c$ through pair-breaking mechanisms. In some TMD systems moderate disorder can weaken competing charge-density-wave order, thereby enhancing superconductivity, whereas stronger disorder ultimately suppresses superconducting pairing.

These observations demonstrate that the superconducting phase diagram depends directly on the defect landscape established during crystal growth.

\begin{figure*}
	\centering
	\includegraphics[width=0.9\textwidth]{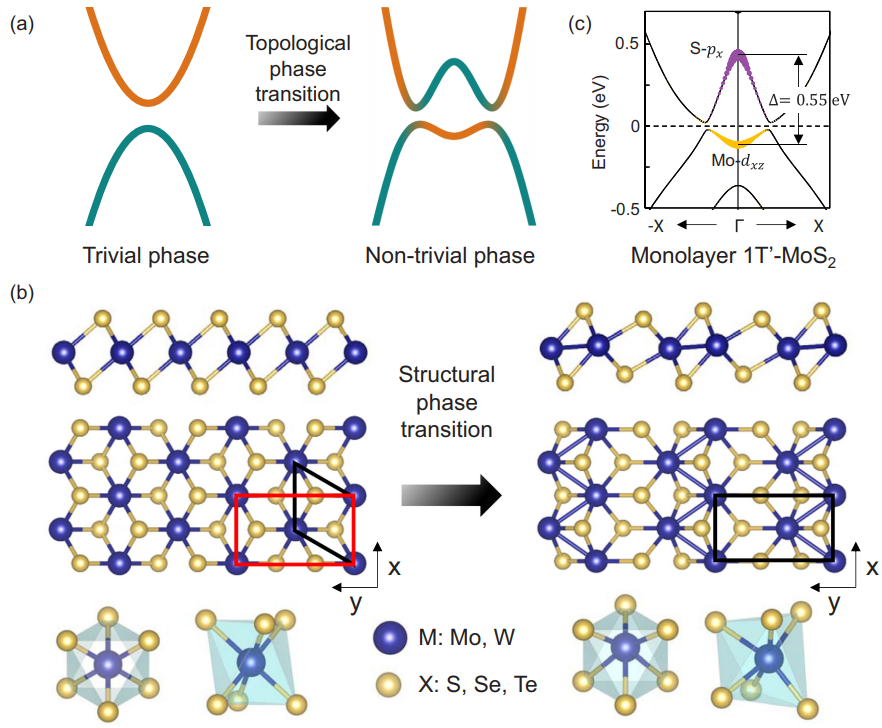}
	\caption{
		\textbf{Topological band inversion and structural phase transformation in transition metal dichalcogenides.}
		(a) Schematic illustration of a topological phase transition in a time-reversal-invariant insulator. The relative ordering of valence and conduction bands reverses, and hybridization between inverted states opens a nontrivial bulk energy gap. 
		(b) Structural transformation from the 1T to the 1T$'$ phase in TMDs. Black solid lines denote the primitive unit cells, while red solid lines indicate the rectangular supercell associated with the distorted structure. 
		(c) Calculated band structure of monolayer 1T$'$-MoS$_2$ near the $\Gamma$ point, exhibiting an apparent inverted gap of approximately $\Delta \approx 0.55$\,eV. 
		Reproduced with permission from Ref.~\cite{Choe_2016}.
	}
	\label{fig:topo}
\end{figure*}

\subsection{Topological Phases and Structural Symmetry}

In layered tellurides such as MoTe$_2$ and WTe$_2$, electronic topology depends sensitively on lattice symmetry and structural distortion~\cite{Qian_2014,Ali_2014}. The distinction between the 1T$'$ and T$_\mathrm{d}$ structures involves subtle stacking changes that determine inversion symmetry and the presence of Weyl nodes.

Figure~\ref{fig:topo} highlights the connection between structural distortion and band inversion~\cite{Choe_2016}. Because the energy difference between competing polytypes is small, growth conditions that stabilize one structure over another determine whether a topologically nontrivial band structure is realized. Strain, stoichiometry, and cooling rate therefore play important roles in structural selection.

A representative example occurs in MoTe$_2$, where the energy difference between the centrosymmetric $1T'$ phase and the noncentrosymmetric $T_d$ phase is only $\Delta E \sim 20$--$40$\,meV per formula unit~\cite{Qian_2014,Soluyanov_2015}. Because this energy scale is comparable to thermal energies during synthesis, modest variations in cooling rate or stoichiometry can determine which phase is retained. Rapid quenching can kinetically trap the high-temperature structure, whereas slow cooling favors the thermodynamic ground state. Since inversion symmetry determines whether Weyl nodes appear in the electronic band structure, crystal growth can therefore control whether the material realizes a trivial or topological semimetallic state.

Stoichiometric deviations can also modify electronic topology indirectly through shifts of the Fermi level relative to band crossings. In many topological semimetals the Weyl or Dirac nodes occur at specific energies close to the Fermi level. Defects introduced during growth, such as chalcogen vacancies, can donate carriers and shift the chemical potential. Even small deviations from stoichiometry can therefore move the Fermi level away from the band-crossing points, altering observable transport signatures.

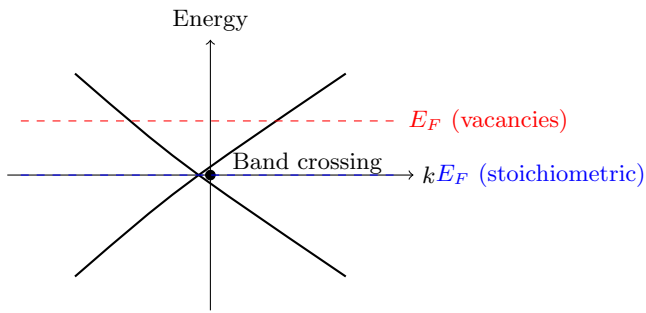
\begin{figure}[h]
	\centering
	\begin{tikzpicture}[scale=0.9]
		
		\draw[->] (-3,0) -- (3,0) node[right] {$k$};
		\draw[->] (0,-2) -- (0,2) node[above] {Energy};
		
		\draw[thick] (-2,-1.5) .. controls (-0.5,-0.2) .. (2,1.5);
		\draw[thick] (-2,1.5) .. controls (-0.5,0.2) .. (2,-1.5);
		
		\filldraw (0,0) circle (2pt);
		\node[above right] at (0.2,-0.1) {Band crossing};
		
		\draw[dashed,blue] (-2.8,0) -- (2.8,0);
		\node[right,blue] at (3.2,0) {$E_F$ (stoichiometric)};
		
		\draw[dashed,red] (-2.8,0.8) -- (2.8,0.8);
		\node[right,red] at (2.8,0.8) {$E_F$ (vacancies)};
		
	\end{tikzpicture}
	
	\caption{
		Schematic illustration of Fermi-level shifts induced by stoichiometric deviations in topological semimetals. In the ideal stoichiometric case (blue dashed line), the Fermi level lies near the band-crossing point. Chalcogen vacancies can donate carriers and shift the Fermi level upward (red dashed line), moving it away from the band crossing and modifying observable transport signatures while the underlying band topology remains unchanged.
	}
	\label{fig:fermi_shift_topology}
\end{figure}

\subsection{Correlation Effects and Mott Physics}

Layered compounds such as 1T-TaS$_2$ exhibit correlated insulating states associated with commensurate CDW order and electron localization~\cite{Wilson_1975}. The stability of these phases depends on lattice distortion, electronic filling, and electronic screening. Defects introduced during synthesis modify local charge density and interaction strength, potentially destabilizing correlated states.

In narrow-bandwidth systems, even dilute disorder can modify hopping amplitudes and exchange pathways. Vacancy concentration and stacking faults therefore influence both the single-particle band structure and many-body interactions. Representative examples illustrating these relationships are summarized in Table~\ref{tab:growth_phase_mapping}. The table highlights how specific synthesis conditions, including chalcogen chemical potential, cooling rate, and growth disorder, produce distinct structural modifications such as vacancy formation, polytype stabilization, or grain-boundary networks. These structural variations directly influence measurable physical properties including carrier density, mobility, superconducting transition temperature, and charge-density-wave coherence.

\begin{table*}[htbp]
	\centering
	\caption{Representative examples illustrating the connection between growth conditions, structural modification, and resulting physical properties in transition metal dichalcogenides.}
	\label{tab:growth_phase_mapping}
	
	\begin{tblr}{
			colspec = {p{2.5cm} p{5.0cm} p{5.0cm} p{4.0cm}},
			row{1} = {font=\bfseries},
			hlines,
			cells = {t},             
			abovesep = 4pt,
			belowsep = 4pt,
			stretch = 1.5,           
		}
		Growth Variable & Structural / Defect Change & Physical Consequence & Representative System \\
		\hline
		Chalcogen-poor conditions & 
		Increased chalcogen vacancies & 
		Enhanced carrier density and reduced mobility & 
		MoS$_2$~\cite{Jaques_2026} \\
		High residual disorder & 
		Shortened CDW coherence length & 
		Broadened CDW transition and modified $T_c$ & 
		NbSe$_2$~\cite{Cho_2018} \\
		Stoichiometric deviation & 
		Shifted carrier concentration & 
		Suppressed or shifted CDW transition & 
		TiSe$_2$~\cite{Saini_2023} \\
		Polytype stabilization & 
		1T$'$ vs.\ T$_\mathrm{d}$ selection & 
		Modified band topology and magnetoresistance & 
		MoTe$_2$, WTe$_2$~\cite{Paul_2023} \\
		Rapid cooling & 
		Metastable phase retention & 
		Altered symmetry and electronic structure & 
		Telluride TMDs~\cite{Sokolikova_2020} \\
		Grain-boundary density & 
		Extended defect networks & 
		Increased scattering and reduced mobility & 
		CVD-grown monolayers~\cite{Yu_2011} \\
		\hline
	\end{tblr}
\end{table*}

\subsection{Quantitative Impact of Growth-Induced Disorder}

Quantitative studies demonstrate measurable shifts in collective transition temperatures and transport coefficients arising from growth-dependent disorder. For a point defect with formation energy $E_f$, the equilibrium concentration scales as $n \propto \exp(-E_f/k_B T)$. First-principles calculations for monolayer MoS$_2$ indicate that sulfur vacancy formation energies vary from approximately $E_f \approx 1.5$\,eV under sulfur-rich conditions to $E_f \approx 1.0$\,eV under sulfur-poor conditions~\cite{Komsa_2015}. At a typical synthesis temperature of $T \sim 1000$\,K, this change of $\sim 0.5$\,eV increases the equilibrium vacancy concentration by roughly three orders of magnitude.

Such variations directly modify carrier density and scattering rates in semiconducting TMDs. Comparable growth-dependent effects are observed in metallic systems. In 2H-NbSe$_2$, differences in crystal quality reflected in the residual resistivity ratio correlate with measurable shifts of the superconducting transition temperature $T_c$ by several tenths of a kelvin~\cite{Wilson_1975}. These examples illustrate how modest variations in chemical potentials and defect formation energies during growth propagate directly into measurable electronic properties.

While the preceding discussion emphasizes general thermodynamic and kinetic principles, experimental studies increasingly demonstrate direct correlations between specific synthesis parameters and measurable electronic properties. Examples include variations in chalcogen chemical potential during growth, temperature gradients in vapor transport methods, or precursor flux ratios in vapor-phase deposition. These parameters influence defect densities, stoichiometry, and structural phase selection, which in turn modify carrier concentration, scattering rates, and collective electronic ordering. Representative examples are summarized in Table~\ref{tab:synthesis_property_map}.

\begin{table*}[htbp]
	\centering
	\caption{Examples of experimentally observed correlations between synthesis parameters and electronic properties in transition metal dichalcogenides}
	\label{tab:synthesis_property_map}
	
	\begin{tblr}{
			colspec = {p{2.0cm} p{3.5cm} p{3.5cm} p{5.0cm} p{2.0cm}},
			row{1} = {font=\bfseries},
			hlines,
			cells = {t},             
			abovesep = 4pt,
			belowsep = 4pt,
			stretch = 1.5,           
		}
		Material & Growth parameter & Structural / defect change & Electronic consequence & Reference \\
		\hline
		MoS$_2$ & 
		Sulfur-poor growth conditions & 
		Increased sulfur vacancy density & 
		Enhanced electron doping and reduced carrier mobility due to defect scattering & 
		\cite{Komsa_2015,Manzeli_2017} \\
		NbSe$_2$ & 
		Variation in crystal quality (RRR $\sim$10--40) & 
		Reduced disorder and impurity scattering & 
		$T_c$ shifts by $\sim$0.2--0.4 K and sharper CDW transition & 
		\cite{Wilson_1975,Cho_2018} \\
		TiSe$_2$ & 
		Stoichiometry variation during growth & 
		Change in carrier concentration and defect density & 
		Suppression or shift of the CDW transition temperature & 
		\cite{DiSalvo_1976,Saini_2023} \\
		MoTe$_2$ & 
		Cooling rate and thermal history & 
		Stabilization of $1T'$ or $T_d$ polytype & 
		Transition between trivial and Weyl semimetal electronic states & 
		\cite{Qian_2014,Soluyanov_2015} \\
		CVD-grown MoS$_2$ & 
		Precursor flux ratio and substrate temperature & 
		Variation in grain-boundary density and domain size & 
		Reduced carrier mobility and enhanced scattering at grain boundaries & 
		\cite{Lee_2012,VanDerZande_2013} \\
		\hline
	\end{tblr}
\end{table*}

\subsection{Growth as a Control Parameter for the Effective Hamiltonian}

The conceptual relationships between synthesis conditions, structural outcomes, and electronic properties can be summarized within a unified framework, as illustrated in Table~\ref{tab:growth_structure_electronics}. Variations in growth parameters, including chemical potentials, temperature gradients, precursor flux ratios, and cooling rates, modify the structural and defect landscape through mechanisms such as vacancy formation, grain-boundary generation, and polytype stabilization. These structural changes propagate directly to electronic observables by altering carrier density, scattering rates, lattice symmetry, and band topology. 

This mapping highlights the central theme of this Review: crystal growth establishes the thermodynamic and kinetic boundary conditions that determine the effective electronic Hamiltonian realized in experiment.

\begin{table*}[htbp]
	\centering
	\caption{Schematic mapping between growth parameters, structural modifications, and resulting electronic properties in transition metal dichalcogenides}
	\label{tab:growth_structure_electronics}
	
	\begin{tblr}{
			colspec = {p{3.0cm} p{5.0cm} p{5.0cm} p{3.5cm}},
			row{1} = {font=\bfseries},
			hlines,
			cells = {t},             
			abovesep = 4pt,
			belowsep = 4pt,
			stretch = 1.5,           
		}
		Growth parameter & Structural / defect outcome & Electronic consequence & Representative system \\
		\hline
		Chalcogen chemical potential ($\mu_X$) & 
		Variation in chalcogen vacancy concentration and antisite defects & 
		Carrier density modulation, defect scattering, and mobility variation & 
		MoS$_2$, WS$_2$~\cite{Komsa_2015,Manzeli_2017} \\
		Temperature gradient (CVT / PVT) & 
		Spatial variation in supersaturation and stoichiometry & 
		Changes in defect density, residual resistivity ratio, and superconducting $T_c$ & 
		NbSe$_2$~\cite{Wilson_1975,Cho_2018} \\
		Cooling rate & 
		Retention of metastable polytypes or structural distortions & 
		Modification of lattice symmetry and possible realization of topological phases & 
		MoTe$_2$, WTe$_2$~\cite{Qian_2014,Soluyanov_2015} \\
		Precursor flux ratio (thin films) & 
		Variation in nucleation density and grain-boundary networks & 
		Enhanced carrier scattering and reduced carrier mobility & 
		CVD-grown MoS$_2$~\cite{Lee_2012,VanDerZande_2013} \\
		Substrate-induced strain & 
		Stabilization of competing polymorphs and lattice distortions & 
		Band structure modification and possible changes in electronic topology & 
		MoTe$_2$, WTe$_2$~\cite{Qian_2014,Ahn_2017} \\
		Growth disorder / impurity incorporation & 
		Point defects and impurity scattering centers & 
		Modification of CDW coherence length and superconducting transition temperature & 
		NbSe$_2$, TiSe$_2$~\cite{Wilson_1975,DiSalvo_1976} \\
		\hline
	\end{tblr}
\end{table*}

Crystal growth therefore defines the boundary conditions under which emergent quantum phases appear. Even small variations in chemical potentials during synthesis can produce large changes in defect density and measurable variations in macroscopic electronic behavior. A complete understanding of TMD physics therefore requires explicit consideration of synthesis history alongside electronic structure.


\section{Emerging Frontiers and Deterministic Control}

The preceding sections establish that crystal growth defines chemical potentials, defect populations, and structural symmetry that ultimately determine electronic behavior in transition metal dichalcogenides. A natural next step is the pursuit of deterministic control over these parameters. Rather than relying primarily on empirical optimization, future synthesis strategies can be guided by explicit thermodynamic constraints and kinetic rate equations. From this perspective, crystal growth evolves from a largely empirical practice toward a predictive tool for materials design.

\subsection{Deterministic Defect Engineering}

Equilibrium defect concentrations scale exponentially with defect formation energies. Because these formation energies depend on $\mu_M$, $\mu_X$, and temperature, controlled tuning of chemical potentials provides a direct route to tailoring intrinsic doping levels and scattering rates~\cite{Zunger_2021}.

In practice, maintaining reproducible chemical potentials remains challenging, particularly in vapor-phase techniques where transport-agent reactions, partial pressures, and temperature gradients vary spatially. Progress will require quantitative calibration of chemical potential boundaries under realistic growth conditions. In thin-film methods such as molecular beam epitaxy, independent control of metal and chalcogen fluxes allows systematic exploration of chalcogen-rich and chalcogen-poor regimes. Coupling first-principles defect thermodynamics with calibrated flux control could enable deliberate tuning of vacancy concentrations and carrier density.

Such approaches would allow controlled mapping between defect density and measurable observables including carrier mobility, optical linewidth, superconducting transition temperature, and CDW coherence length.

\subsection{Nonequilibrium Phase Stabilization}

Many TMD systems exhibit free-energy differences between competing polytypes on the order of tens of meV per formula unit. Under these conditions, kinetic barriers rather than equilibrium thermodynamics may determine the realized structure. Rapid cooling, pulsed precursor delivery, or strain-assisted growth can stabilize metastable phases that are not favored in equilibrium phase diagrams~\cite{Shuang_2026}.

A quantitative understanding of activation barriers between polytypes remains limited. Measurements of transformation kinetics, combined with atomistic modeling of diffusion and lattice rearrangement, are required to determine the temperature and time scales governing phase interconversion. Establishing these scales would enable intentional stabilization of structures with targeted symmetry or electronic topology.

\subsection{Strain and Interface Engineering}

In thin films and heterostructures, interface and strain energies enter directly into the free-energy balance. Even weak lattice mismatch can shift the relative stability of competing polymorphs when intrinsic energy differences are small~\cite{Ahn_2017}. Controlled strain through lattice-mismatched substrates, flexible platforms, or patterned confinement therefore provides an additional thermodynamic control parameter.

Interface engineering is equally important in vertical heterostructures. Clean interfaces and minimized interdiffusion are required to investigate proximity effects and interlayer coupling. The thermodynamic--kinetic framework developed earlier can guide the selection of temperature and flux regimes that minimize unwanted reactions while preserving structural coherence.

\subsection{In Situ Diagnostics and Feedback Control}

A major limitation in current TMD synthesis is incomplete real-time knowledge of evolving chemical potentials and defect incorporation. In thin-film growth, techniques such as reflection high-energy electron diffraction, optical spectroscopy, and mass spectrometry provide partial insight into surface structure and vapor composition. Comparable diagnostic capabilities remain less developed for bulk growth methods.

Integrating in situ monitoring with feedback control of temperature gradients or precursor flux could significantly reduce batch-to-batch variability. Coupling thermodynamic modeling with experimental monitoring would allow dynamic adjustment of supersaturation and chemical potentials to maintain growth within defined stability windows. Recent advances in autonomous synthesis platforms and machine-learning-guided materials discovery have enabled closed-loop experimental workflows in which algorithms iteratively propose, execute, and analyze experiments to optimize synthesis conditions and accelerate materials discovery~\cite{Stach_2021,Kusne_2020,Szymanski_2023}.

\subsection{Scalability and Reproducibility}

For device integration, scalable growth with uniform defect density and phase purity is essential. Variability in grain size, stoichiometry, and vacancy concentration currently limits reproducibility across laboratories. The transport and supersaturation analysis developed in Sections II--IV provides a foundation for rational scaling by identifying which parameters most strongly influence nucleation density and defect incorporation.

Standardized reporting of temperature gradients, precursor ratios, transport-agent concentration, cooling rates, and post-growth treatments would improve cross-study comparability. Reproducibility ultimately depends on explicit documentation of the thermodynamic and kinetic conditions under which a given phase is stabilized.

\subsection{Open Questions}

Several central questions remain. How precisely can chemical potentials be stabilized during high-temperature vapor growth? What are the quantitative activation barriers governing polytype transformations in key TMD systems? How does reduced dimensionality modify defect formation energies under realistic nonequilibrium growth conditions?

Addressing these questions requires coordinated experimental and theoretical efforts. First-principles calculations must be linked to experimentally calibrated chemical potentials and kinetic rate constants. Only through the integration of thermodynamic modeling, kinetic analysis, and controlled synthesis can crystal growth evolve from an empirical procedure into a predictive platform for engineering functional quantum materials.


\section{Conclusions}

Transition metal dichalcogenides provide a versatile platform for realizing semiconducting, metallic, correlated, and topological electronic states. Across this class of materials, the structural and defect landscape is determined not only by chemical composition but also by the thermodynamic and kinetic conditions under which crystals are synthesized. Chemical potentials define phase and defect stability boundaries, supersaturation governs nucleation density and morphology, and mass-transport regimes determine whether equilibrium configurations are reached or kinetically bypassed during growth.

By combining thermodynamic stability analysis with kinetic growth theory, a unified framework emerges for interpreting synthesis-dependent variability in TMD systems. Bulk techniques such as chemical vapor transport, flux growth, and physical vapor transport access distinct regions of chemical potential and supersaturation space, leading to characteristic defect populations and structural outcomes. Thin-film methods introduce additional contributions from two-dimensional nucleation, interfacial energy, and strain, which modify the effective free-energy landscape. In all cases, growth conditions determine stoichiometry, vacancy concentration, stacking order, and microstructure.

These structural variables directly enter the effective electronic Hamiltonian. Charge-density-wave coherence, superconducting transition temperature, carrier density, and band topology respond sensitively to defect concentration and phase selection~\cite{Sung_2024,Li_2021_2}. Consequently, variations in reported physical properties across nominally identical compounds frequently originate from differences in chemical potential control, cooling history, or transport regime rather than intrinsic discrepancies in electronic structure.

The central conclusion of this Review is that crystal growth establishes the thermodynamic and kinetic boundary conditions under which emergent quantum phases are realized. Treating synthesis as an integral component of materials physics, rather than merely a preparative step, clarifies the origin of sample-dependent behavior and provides a pathway toward reproducible phase control. A quantitative integration of thermodynamic modeling, kinetic analysis, and experimentally calibrated growth protocols will be essential for systematic navigation of phase space in transition metal dichalcogenides and related layered quantum materials. Future advances in predictive synthesis, autonomous experimentation, and in situ diagnostics may further enable deterministic control of structure and emergent quantum phases in layered materials.

\section*{Author contributions}
Anzar Ali: Conceptualization, Investigation, Writing - original draft.
Md Ezaz Hasan Khan: Supervision, Writing - review \& editing, Funding acquisition.
Mahmoud Abdel-Hafiez: Conceptualization, Supervision, Writing - review \& editing, Funding acquisition.

\section*{Conflicts of interest}

The authors declare that there are no conflicts of interest.

\section*{Data availability}

No new data were created or analysed in this study. Data sharing is not applicable to this article.

\section*{Acknowledgments}

Research reported in this publication was supported by the Qatar Research Development and Innovation Council (QRDI) under Project No.~ARG01-0516-230179. The authors also gratefully acknowledge support from the University of Doha for Science and Technology, Qatar.


\begin{thebibliography}{61}%
	\makeatletter
	\providecommand \@ifxundefined [1]{%
		\@ifx{#1\undefined}
	}%
	\providecommand \@ifnum [1]{%
		\ifnum #1\expandafter \@firstoftwo
		\else \expandafter \@secondoftwo
		\fi
	}%
	\providecommand \@ifx [1]{%
		\ifx #1\expandafter \@firstoftwo
		\else \expandafter \@secondoftwo
		\fi
	}%
	\providecommand \natexlab [1]{#1}%
	\providecommand \enquote  [1]{``#1''}%
	\providecommand \bibnamefont  [1]{#1}%
	\providecommand \bibfnamefont [1]{#1}%
	\providecommand \citenamefont [1]{#1}%
	\providecommand \href@noop [0]{\@secondoftwo}%
	\providecommand \href [0]{\begingroup \@sanitize@url \@href}%
	\providecommand \@href[1]{\@@startlink{#1}\@@href}%
	\providecommand \@@href[1]{\endgroup#1\@@endlink}%
	\providecommand \@sanitize@url [0]{\catcode `\\12\catcode `\$12\catcode
		`\&12\catcode `\#12\catcode `\^12\catcode `\_12\catcode `\%12\relax}%
	\providecommand \@@startlink[1]{}%
	\providecommand \@@endlink[0]{}%
	\providecommand \url  [0]{\begingroup\@sanitize@url \@url }%
	\providecommand \@url [1]{\endgroup\@href {#1}{\urlprefix }}%
	\providecommand \urlprefix  [0]{URL }%
	\providecommand \Eprint [0]{\href }%
	\providecommand \doibase [0]{https://doi.org/}%
	\providecommand \selectlanguage [0]{\@gobble}%
	\providecommand \bibinfo  [0]{\@secondoftwo}%
	\providecommand \bibfield  [0]{\@secondoftwo}%
	\providecommand \translation [1]{[#1]}%
	\providecommand \BibitemOpen [0]{}%
	\providecommand \bibitemStop [0]{}%
	\providecommand \bibitemNoStop [0]{.\EOS\space}%
	\providecommand \EOS [0]{\spacefactor3000\relax}%
	\providecommand \BibitemShut  [1]{\csname bibitem#1\endcsname}%
	\let\auto@bib@innerbib\@empty
	\bibitem [{\citenamefont {Manzeli}\ \emph {et~al.}(2017)\citenamefont
		{Manzeli}, \citenamefont {Ovchinnikov}, \citenamefont {Pasquier},
		\citenamefont {Yazyev},\ and\ \citenamefont {Kis}}]{Manzeli_2017}%
	\BibitemOpen
	\bibfield  {author} {\bibinfo {author} {\bibfnamefont {S.}~\bibnamefont
			{Manzeli}}, \bibinfo {author} {\bibfnamefont {D.}~\bibnamefont
			{Ovchinnikov}}, \bibinfo {author} {\bibfnamefont {D.}~\bibnamefont
			{Pasquier}}, \bibinfo {author} {\bibfnamefont {O.~V.}\ \bibnamefont
			{Yazyev}},\ and\ \bibinfo {author} {\bibfnamefont {A.}~\bibnamefont {Kis}},\
	}\bibfield  {title} {\bibinfo {title} {{2D Transition Metal
				Dichalcogenides}},\ }\href {https://doi.org/10.1038/natrevmats.2017.33}
	{\bibfield  {journal} {\bibinfo  {journal} {Nat. Rev. Mater.}\ }\textbf
		{\bibinfo {volume} {2}},\ \bibinfo {pages} {17033} (\bibinfo {year}
		{2017})}\BibitemShut {NoStop}%
	\bibitem [{\citenamefont {Wilson}\ \emph {et~al.}(1975)\citenamefont {Wilson},
		\citenamefont {Di~Salvo},\ and\ \citenamefont {Mahajan}}]{Wilson_1975}%
	\BibitemOpen
	\bibfield  {author} {\bibinfo {author} {\bibfnamefont {J.~A.}\ \bibnamefont
			{Wilson}}, \bibinfo {author} {\bibfnamefont {F.~J.}\ \bibnamefont
			{Di~Salvo}},\ and\ \bibinfo {author} {\bibfnamefont {S.}~\bibnamefont
			{Mahajan}},\ }\bibfield  {title} {\bibinfo {title} {{Charge-Density Waves and
				Superlattices in Metallic Layered Transition Metal Dichalcogenides}},\ }\href
	{https://doi.org/10.1080/00018737500101391} {\bibfield  {journal} {\bibinfo
			{journal} {Adv. Phys.}\ }\textbf {\bibinfo {volume} {24}},\ \bibinfo {pages}
		{117} (\bibinfo {year} {1975})}\BibitemShut {NoStop}%
	\bibitem [{\citenamefont {Mak}\ \emph {et~al.}(2010)\citenamefont {Mak},
		\citenamefont {Lee}, \citenamefont {Hone}, \citenamefont {Shan},\ and\
		\citenamefont {Heinz}}]{Mak_2010}%
	\BibitemOpen
	\bibfield  {author} {\bibinfo {author} {\bibfnamefont {K.~F.}\ \bibnamefont
			{Mak}}, \bibinfo {author} {\bibfnamefont {C.}~\bibnamefont {Lee}}, \bibinfo
		{author} {\bibfnamefont {J.}~\bibnamefont {Hone}}, \bibinfo {author}
		{\bibfnamefont {J.}~\bibnamefont {Shan}},\ and\ \bibinfo {author}
		{\bibfnamefont {T.~F.}\ \bibnamefont {Heinz}},\ }\bibfield  {title} {\bibinfo
		{title} {{Atomically Thin {MoS$_2$}: A New Direct-Gap Semiconductor}},\
	}\href {https://doi.org/10.1103/PhysRevLett.105.136805} {\bibfield  {journal}
		{\bibinfo  {journal} {Phys. Rev. Lett.}\ }\textbf {\bibinfo {volume} {105}},\
		\bibinfo {pages} {136805} (\bibinfo {year} {2010})}\BibitemShut {NoStop}%
	\bibitem [{\citenamefont {Splendiani}\ \emph {et~al.}(2010)\citenamefont
		{Splendiani}, \citenamefont {Sun}, \citenamefont {Zhang}, \citenamefont {Li},
		\citenamefont {Kim}, \citenamefont {Galli},\ and\ \citenamefont
		{Wang}}]{Splendiani_2010}%
	\BibitemOpen
	\bibfield  {author} {\bibinfo {author} {\bibfnamefont {A.}~\bibnamefont
			{Splendiani}}, \bibinfo {author} {\bibfnamefont {L.}~\bibnamefont {Sun}},
		\bibinfo {author} {\bibfnamefont {Y.}~\bibnamefont {Zhang}}, \bibinfo
		{author} {\bibfnamefont {T.}~\bibnamefont {Li}}, \bibinfo {author}
		{\bibfnamefont {J.}~\bibnamefont {Kim}}, \bibinfo {author} {\bibfnamefont
			{G.}~\bibnamefont {Galli}},\ and\ \bibinfo {author} {\bibfnamefont
			{F.}~\bibnamefont {Wang}},\ }\bibfield  {title} {\bibinfo {title} {{Emerging
				Photoluminescence in Monolayer {MoS$_2$}}},\ }\href
	{https://doi.org/10.1021/nl903868w} {\bibfield  {journal} {\bibinfo
			{journal} {Nano Lett.}\ }\textbf {\bibinfo {volume} {10}},\ \bibinfo {pages}
		{1271} (\bibinfo {year} {2010})}\BibitemShut {NoStop}%
	\bibitem [{\citenamefont {Moncton}\ \emph {et~al.}(1975)\citenamefont
		{Moncton}, \citenamefont {Axe},\ and\ \citenamefont
		{DiSalvo}}]{Moncton_1975}%
	\BibitemOpen
	\bibfield  {author} {\bibinfo {author} {\bibfnamefont {D.~E.}\ \bibnamefont
			{Moncton}}, \bibinfo {author} {\bibfnamefont {J.~D.}\ \bibnamefont {Axe}},\
		and\ \bibinfo {author} {\bibfnamefont {F.~J.}\ \bibnamefont {DiSalvo}},\
	}\bibfield  {title} {\bibinfo {title} {{Study of superlattice formation in
				$2H$-NbSe$_2$ and $2H$-TaSe$_2$ by neutron scattering}},\ }\href
	{https://doi.org/10.1103/PhysRevLett.34.734} {\bibfield  {journal} {\bibinfo
			{journal} {Phys. Rev. Lett.}\ }\textbf {\bibinfo {volume} {34}},\ \bibinfo
		{pages} {734} (\bibinfo {year} {1975})}\BibitemShut {NoStop}%
	\bibitem [{\citenamefont {Di~Salvo}\ \emph {et~al.}(1976)\citenamefont
		{Di~Salvo}, \citenamefont {Moncton},\ and\ \citenamefont
		{Waszczak}}]{DiSalvo_1976}%
	\BibitemOpen
	\bibfield  {author} {\bibinfo {author} {\bibfnamefont {F.~J.}\ \bibnamefont
			{Di~Salvo}}, \bibinfo {author} {\bibfnamefont {D.~E.}\ \bibnamefont
			{Moncton}},\ and\ \bibinfo {author} {\bibfnamefont {J.~V.}\ \bibnamefont
			{Waszczak}},\ }\bibfield  {title} {\bibinfo {title} {{Electronic properties
				and superlattice formation in the semimetal ${\mathrm{TiSe}}_{2}$}},\ }\href
	{https://doi.org/10.1103/PhysRevB.14.4321} {\bibfield  {journal} {\bibinfo
			{journal} {Phys. Rev. B}\ }\textbf {\bibinfo {volume} {14}},\ \bibinfo
		{pages} {4321} (\bibinfo {year} {1976})}\BibitemShut {NoStop}%
	\bibitem [{\citenamefont {Fazekas}\ and\ \citenamefont
		{Tosatti}(1979)}]{Fazekas_1979}%
	\BibitemOpen
	\bibfield  {author} {\bibinfo {author} {\bibfnamefont {P.}~\bibnamefont
			{Fazekas}}\ and\ \bibinfo {author} {\bibfnamefont {E.}~\bibnamefont
			{Tosatti}},\ }\bibfield  {title} {\bibinfo {title} {{Electrical, structural
				and magnetic properties of pure and doped 1T-TaS$_2$}},\ }\href
	{https://doi.org/10.1080/13642817908245359} {\bibfield  {journal} {\bibinfo
			{journal} {Philosophical Magazine B}\ }\textbf {\bibinfo {volume} {39}},\
		\bibinfo {pages} {229} (\bibinfo {year} {1979})}\BibitemShut {NoStop}%
	\bibitem [{\citenamefont {Ali}\ \emph {et~al.}(2014)\citenamefont {Ali},
		\citenamefont {Xiong}, \citenamefont {Flynn}, \citenamefont {Tao},
		\citenamefont {Gibson}, \citenamefont {Schoop}, \citenamefont {Liang},
		\citenamefont {Haldolaarachchige}, \citenamefont {Hirschberger},
		\citenamefont {Ong},\ and\ \citenamefont {Cava}}]{Ali_2014}%
	\BibitemOpen
	\bibfield  {author} {\bibinfo {author} {\bibfnamefont {M.~N.}\ \bibnamefont
			{Ali}}, \bibinfo {author} {\bibfnamefont {J.}~\bibnamefont {Xiong}}, \bibinfo
		{author} {\bibfnamefont {S.}~\bibnamefont {Flynn}}, \bibinfo {author}
		{\bibfnamefont {J.}~\bibnamefont {Tao}}, \bibinfo {author} {\bibfnamefont
			{Q.~D.}\ \bibnamefont {Gibson}}, \bibinfo {author} {\bibfnamefont {L.~M.}\
			\bibnamefont {Schoop}}, \bibinfo {author} {\bibfnamefont {T.}~\bibnamefont
			{Liang}}, \bibinfo {author} {\bibfnamefont {N.}~\bibnamefont
			{Haldolaarachchige}}, \bibinfo {author} {\bibfnamefont {M.}~\bibnamefont
			{Hirschberger}}, \bibinfo {author} {\bibfnamefont {N.~P.}\ \bibnamefont
			{Ong}},\ and\ \bibinfo {author} {\bibfnamefont {R.~J.}\ \bibnamefont
			{Cava}},\ }\bibfield  {title} {\bibinfo {title} {{Large, Non-Saturating
				Magnetoresistance in {WTe$_2$}}},\ }\href
	{https://doi.org/10.1038/nature13763} {\bibfield  {journal} {\bibinfo
			{journal} {Nature}\ }\textbf {\bibinfo {volume} {514}},\ \bibinfo {pages}
		{205} (\bibinfo {year} {2014})}\BibitemShut {NoStop}%
	\bibitem [{\citenamefont {Qian}\ \emph {et~al.}(2014)\citenamefont {Qian},
		\citenamefont {Liu}, \citenamefont {Fu},\ and\ \citenamefont
		{Li}}]{Qian_2014}%
	\BibitemOpen
	\bibfield  {author} {\bibinfo {author} {\bibfnamefont {X.}~\bibnamefont
			{Qian}}, \bibinfo {author} {\bibfnamefont {J.}~\bibnamefont {Liu}}, \bibinfo
		{author} {\bibfnamefont {L.}~\bibnamefont {Fu}},\ and\ \bibinfo {author}
		{\bibfnamefont {J.}~\bibnamefont {Li}},\ }\bibfield  {title} {\bibinfo
		{title} {{Quantum spin Hall effect in two-dimensional transition metal
				dichalcogenides}},\ }\href {https://doi.org/10.1126/science.1256815}
	{\bibfield  {journal} {\bibinfo  {journal} {Science}\ }\textbf {\bibinfo
			{volume} {346}},\ \bibinfo {pages} {1344} (\bibinfo {year}
		{2014})}\BibitemShut {NoStop}%
	\bibitem [{\citenamefont {Soluyanov}\ \emph {et~al.}(2015)\citenamefont
		{Soluyanov}, \citenamefont {Gresch}, \citenamefont {Wang}, \citenamefont
		{Wu}, \citenamefont {Troyer}, \citenamefont {Dai},\ and\ \citenamefont
		{Bernevig}}]{Soluyanov_2015}%
	\BibitemOpen
	\bibfield  {author} {\bibinfo {author} {\bibfnamefont {A.~A.}\ \bibnamefont
			{Soluyanov}}, \bibinfo {author} {\bibfnamefont {D.}~\bibnamefont {Gresch}},
		\bibinfo {author} {\bibfnamefont {Z.}~\bibnamefont {Wang}}, \bibinfo {author}
		{\bibfnamefont {Q.}~\bibnamefont {Wu}}, \bibinfo {author} {\bibfnamefont
			{M.}~\bibnamefont {Troyer}}, \bibinfo {author} {\bibfnamefont
			{X.}~\bibnamefont {Dai}},\ and\ \bibinfo {author} {\bibfnamefont {B.~A.}\
			\bibnamefont {Bernevig}},\ }\bibfield  {title} {\bibinfo {title} {{Type-II
				Weyl semimetals}},\ }\href {https://doi.org/10.1038/nature15768} {\bibfield
		{journal} {\bibinfo  {journal} {Nature}\ }\textbf {\bibinfo {volume} {527}},\
		\bibinfo {pages} {495} (\bibinfo {year} {2015})}\BibitemShut {NoStop}%
	\bibitem [{\citenamefont {Chen}\ \emph {et~al.}(2022)\citenamefont {Chen},
		\citenamefont {Zhang}, \citenamefont {Kan}, \citenamefont {He}, \citenamefont
		{Song}, \citenamefont {Pang}, \citenamefont {Wei},\ and\ \citenamefont
		{Chen}}]{Chen_2022}%
	\BibitemOpen
	\bibfield  {author} {\bibinfo {author} {\bibfnamefont {H.}~\bibnamefont
			{Chen}}, \bibinfo {author} {\bibfnamefont {J.}~\bibnamefont {Zhang}},
		\bibinfo {author} {\bibfnamefont {D.}~\bibnamefont {Kan}}, \bibinfo {author}
		{\bibfnamefont {J.}~\bibnamefont {He}}, \bibinfo {author} {\bibfnamefont
			{M.}~\bibnamefont {Song}}, \bibinfo {author} {\bibfnamefont {J.}~\bibnamefont
			{Pang}}, \bibinfo {author} {\bibfnamefont {S.}~\bibnamefont {Wei}},\ and\
		\bibinfo {author} {\bibfnamefont {K.}~\bibnamefont {Chen}},\ }\bibfield
	{title} {\bibinfo {title} {{The Recent Progress of Two-Dimensional Transition
				Metal Dichalcogenides and Their Phase Transition}},\ }\href
	{https://www.mdpi.com/2073-4352/12/10/1381} {\bibfield  {journal} {\bibinfo
			{journal} {Crystals}\ }\textbf {\bibinfo {volume} {12}},\ \bibinfo {pages}
		{1381} (\bibinfo {year} {2022})}\BibitemShut {NoStop}%
	\bibitem [{\citenamefont {Zhang}\ and\ \citenamefont
		{Northrup}(1991)}]{Zhang_1991}%
	\BibitemOpen
	\bibfield  {author} {\bibinfo {author} {\bibfnamefont {S.~B.}\ \bibnamefont
			{Zhang}}\ and\ \bibinfo {author} {\bibfnamefont {J.~E.}\ \bibnamefont
			{Northrup}},\ }\bibfield  {title} {\bibinfo {title} {{Chemical potential
				dependence of defect formation energies in GaAs: Application to Ga
				self-diffusion}},\ }\href {https://doi.org/10.1103/PhysRevLett.67.2339}
	{\bibfield  {journal} {\bibinfo  {journal} {Physical Review Letters}\
		}\textbf {\bibinfo {volume} {67}},\ \bibinfo {pages} {2339} (\bibinfo {year}
		{1991})}\BibitemShut {NoStop}%
	\bibitem [{\citenamefont {Van~de Walle}\ and\ \citenamefont
		{Neugebauer}(2004)}]{VanDeWalle_2004}%
	\BibitemOpen
	\bibfield  {author} {\bibinfo {author} {\bibfnamefont {C.~G.}\ \bibnamefont
			{Van~de Walle}}\ and\ \bibinfo {author} {\bibfnamefont {J.}~\bibnamefont
			{Neugebauer}},\ }\bibfield  {title} {\bibinfo {title} {{First-principles
				calculations for defects and impurities: Applications to III-nitrides}},\
	}\href {https://doi.org/10.1063/1.1682673} {\bibfield  {journal} {\bibinfo
			{journal} {Journal of Applied Physics}\ }\textbf {\bibinfo {volume} {95}},\
		\bibinfo {pages} {3851} (\bibinfo {year} {2004})}\BibitemShut {NoStop}%
	\bibitem [{\citenamefont {Komsa}\ and\ \citenamefont
		{Krasheninnikov}(2015)}]{Komsa_2015}%
	\BibitemOpen
	\bibfield  {author} {\bibinfo {author} {\bibfnamefont {H.-P.}\ \bibnamefont
			{Komsa}}\ and\ \bibinfo {author} {\bibfnamefont {A.~V.}\ \bibnamefont
			{Krasheninnikov}},\ }\bibfield  {title} {\bibinfo {title} {{Native defects in
				bulk and monolayer MoS$_2$ from first principles}},\ }\href
	{https://doi.org/10.1103/PhysRevB.91.125304} {\bibfield  {journal} {\bibinfo
			{journal} {Physical Review B}\ }\textbf {\bibinfo {volume} {91}},\ \bibinfo
		{pages} {125304} (\bibinfo {year} {2015})}\BibitemShut {NoStop}%
	\bibitem [{\citenamefont {Hong}\ \emph {et~al.}(2015)\citenamefont {Hong},
		\citenamefont {Hu}, \citenamefont {Probert}, \citenamefont {Li},
		\citenamefont {Lv}, \citenamefont {Yang}, \citenamefont {Gu}, \citenamefont
		{Mao}, \citenamefont {Feng}, \citenamefont {Xie}, \citenamefont {Zhang},
		\citenamefont {Wu}, \citenamefont {Zhang}, \citenamefont {Jin}, \citenamefont
		{Ji}, \citenamefont {Zhang}, \citenamefont {Yuan},\ and\ \citenamefont
		{Zhang}}]{Hong_2015}%
	\BibitemOpen
	\bibfield  {author} {\bibinfo {author} {\bibfnamefont {J.}~\bibnamefont
			{Hong}}, \bibinfo {author} {\bibfnamefont {Z.}~\bibnamefont {Hu}}, \bibinfo
		{author} {\bibfnamefont {M.}~\bibnamefont {Probert}}, \bibinfo {author}
		{\bibfnamefont {K.}~\bibnamefont {Li}}, \bibinfo {author} {\bibfnamefont
			{D.}~\bibnamefont {Lv}}, \bibinfo {author} {\bibfnamefont {X.}~\bibnamefont
			{Yang}}, \bibinfo {author} {\bibfnamefont {L.}~\bibnamefont {Gu}}, \bibinfo
		{author} {\bibfnamefont {N.}~\bibnamefont {Mao}}, \bibinfo {author}
		{\bibfnamefont {Q.}~\bibnamefont {Feng}}, \bibinfo {author} {\bibfnamefont
			{L.}~\bibnamefont {Xie}}, \bibinfo {author} {\bibfnamefont {J.}~\bibnamefont
			{Zhang}}, \bibinfo {author} {\bibfnamefont {D.}~\bibnamefont {Wu}}, \bibinfo
		{author} {\bibfnamefont {Z.}~\bibnamefont {Zhang}}, \bibinfo {author}
		{\bibfnamefont {C.}~\bibnamefont {Jin}}, \bibinfo {author} {\bibfnamefont
			{W.}~\bibnamefont {Ji}}, \bibinfo {author} {\bibfnamefont {X.}~\bibnamefont
			{Zhang}}, \bibinfo {author} {\bibfnamefont {J.}~\bibnamefont {Yuan}},\ and\
		\bibinfo {author} {\bibfnamefont {Z.}~\bibnamefont {Zhang}},\ }\bibfield
	{title} {\bibinfo {title} {{Exploring Atomic Defects in Molybdenum Disulphide
				Monolayers}},\ }\href {https://doi.org/10.1038/ncomms7293} {\bibfield
		{journal} {\bibinfo  {journal} {Nat. Commun.}\ }\textbf {\bibinfo {volume}
			{6}},\ \bibinfo {pages} {6293} (\bibinfo {year} {2015})},\ \bibinfo {note}
	{for thermodynamics of chalcogen vacancy formation and control.}\BibitemShut
	{Stop}%
	\bibitem [{\citenamefont {Novoselov}\ \emph {et~al.}(2004)\citenamefont
		{Novoselov}, \citenamefont {Geim}, \citenamefont {Morozov}, \citenamefont
		{Jiang}, \citenamefont {Zhang}, \citenamefont {Dubonos}, \citenamefont
		{Grigorieva},\ and\ \citenamefont {Firsov}}]{Novoselov_2004}%
	\BibitemOpen
	\bibfield  {author} {\bibinfo {author} {\bibfnamefont {K.~S.}\ \bibnamefont
			{Novoselov}}, \bibinfo {author} {\bibfnamefont {A.~K.}\ \bibnamefont {Geim}},
		\bibinfo {author} {\bibfnamefont {S.~V.}\ \bibnamefont {Morozov}}, \bibinfo
		{author} {\bibfnamefont {D.}~\bibnamefont {Jiang}}, \bibinfo {author}
		{\bibfnamefont {Y.}~\bibnamefont {Zhang}}, \bibinfo {author} {\bibfnamefont
			{S.~V.}\ \bibnamefont {Dubonos}}, \bibinfo {author} {\bibfnamefont {I.~V.}\
			\bibnamefont {Grigorieva}},\ and\ \bibinfo {author} {\bibfnamefont {A.~A.}\
			\bibnamefont {Firsov}},\ }\bibfield  {title} {\bibinfo {title} {{Electric
				Field Effect in Atomically Thin Carbon Films}},\ }\href
	{https://doi.org/10.1126/science.1102896} {\bibfield  {journal} {\bibinfo
			{journal} {Science}\ }\textbf {\bibinfo {volume} {306}},\ \bibinfo {pages}
		{666} (\bibinfo {year} {2004})}\BibitemShut {NoStop}%
	\bibitem [{\citenamefont {Lee}\ \emph {et~al.}(2012)\citenamefont {Lee},
		\citenamefont {Zhang}, \citenamefont {Zhang}, \citenamefont {Chang},
		\citenamefont {Lin}, \citenamefont {Chang}, \citenamefont {Yu}, \citenamefont
		{Wang}, \citenamefont {Chang}, \citenamefont {Li},\ and\ \citenamefont
		{Lin}}]{Lee_2012}%
	\BibitemOpen
	\bibfield  {author} {\bibinfo {author} {\bibfnamefont {Y.-H.}\ \bibnamefont
			{Lee}}, \bibinfo {author} {\bibfnamefont {X.-Q.}\ \bibnamefont {Zhang}},
		\bibinfo {author} {\bibfnamefont {W.}~\bibnamefont {Zhang}}, \bibinfo
		{author} {\bibfnamefont {M.-T.}\ \bibnamefont {Chang}}, \bibinfo {author}
		{\bibfnamefont {C.-T.}\ \bibnamefont {Lin}}, \bibinfo {author} {\bibfnamefont
			{K.-D.}\ \bibnamefont {Chang}}, \bibinfo {author} {\bibfnamefont {Y.-C.}\
			\bibnamefont {Yu}}, \bibinfo {author} {\bibfnamefont {J.~T.-W.}\ \bibnamefont
			{Wang}}, \bibinfo {author} {\bibfnamefont {C.-S.}\ \bibnamefont {Chang}},
		\bibinfo {author} {\bibfnamefont {L.-J.}\ \bibnamefont {Li}},\ and\ \bibinfo
		{author} {\bibfnamefont {T.-W.}\ \bibnamefont {Lin}},\ }\bibfield  {title}
	{\bibinfo {title} {{Synthesis of Large-Area {MoS$_2$} Atomic Layers with
				Chemical Vapor Deposition}},\ }\href {https://doi.org/10.1002/adma.201104798}
	{\bibfield  {journal} {\bibinfo  {journal} {Adv. Mater.}\ }\textbf {\bibinfo
			{volume} {24}},\ \bibinfo {pages} {2320} (\bibinfo {year}
		{2012})}\BibitemShut {NoStop}%
	\bibitem [{\citenamefont {Ugeda}\ \emph {et~al.}(2014)\citenamefont {Ugeda},
		\citenamefont {Bradley}, \citenamefont {Shi}, \citenamefont {da~Jornada},
		\citenamefont {Zhang}, \citenamefont {Qiu}, \citenamefont {Ruan},
		\citenamefont {Mo}, \citenamefont {Hussain}, \citenamefont {Shen},
		\citenamefont {Wang}, \citenamefont {Louie},\ and\ \citenamefont
		{Crommie}}]{Ugeda_2016}%
	\BibitemOpen
	\bibfield  {author} {\bibinfo {author} {\bibfnamefont {M.~M.}\ \bibnamefont
			{Ugeda}}, \bibinfo {author} {\bibfnamefont {A.~J.}\ \bibnamefont {Bradley}},
		\bibinfo {author} {\bibfnamefont {S.-F.}\ \bibnamefont {Shi}}, \bibinfo
		{author} {\bibfnamefont {F.~H.}\ \bibnamefont {da~Jornada}}, \bibinfo
		{author} {\bibfnamefont {Y.}~\bibnamefont {Zhang}}, \bibinfo {author}
		{\bibfnamefont {D.~Y.}\ \bibnamefont {Qiu}}, \bibinfo {author} {\bibfnamefont
			{W.}~\bibnamefont {Ruan}}, \bibinfo {author} {\bibfnamefont {S.-K.}\
			\bibnamefont {Mo}}, \bibinfo {author} {\bibfnamefont {Z.}~\bibnamefont
			{Hussain}}, \bibinfo {author} {\bibfnamefont {Z.-X.}\ \bibnamefont {Shen}},
		\bibinfo {author} {\bibfnamefont {F.}~\bibnamefont {Wang}}, \bibinfo {author}
		{\bibfnamefont {S.~G.}\ \bibnamefont {Louie}},\ and\ \bibinfo {author}
		{\bibfnamefont {M.~F.}\ \bibnamefont {Crommie}},\ }\bibfield  {title}
	{\bibinfo {title} {{Giant bandgap renormalization and excitonic effects in a
				monolayer transition metal dichalcogenide semiconductor}},\ }\href
	{https://doi.org/10.1038/nmat4061} {\bibfield  {journal} {\bibinfo  {journal}
			{Nature Materials}\ }\textbf {\bibinfo {volume} {13}},\ \bibinfo {pages}
		{1091} (\bibinfo {year} {2014})}\BibitemShut {NoStop}%
	\bibitem [{\citenamefont {Huang}\ \emph {et~al.}(2011)\citenamefont {Huang},
		\citenamefont {Ruiz-Vargas}, \citenamefont {van~der Zande}, \citenamefont
		{Whitney}, \citenamefont {Levendorf}, \citenamefont {Kevek}, \citenamefont
		{Garg}, \citenamefont {Alden}, \citenamefont {Hustedt}, \citenamefont {Zhu},
		\citenamefont {Park}, \citenamefont {McEuen},\ and\ \citenamefont
		{Muller}}]{Huang_2011}%
	\BibitemOpen
	\bibfield  {author} {\bibinfo {author} {\bibfnamefont {P.~Y.}\ \bibnamefont
			{Huang}}, \bibinfo {author} {\bibfnamefont {C.~S.}\ \bibnamefont
			{Ruiz-Vargas}}, \bibinfo {author} {\bibfnamefont {A.~M.}\ \bibnamefont
			{van~der Zande}}, \bibinfo {author} {\bibfnamefont {W.~S.}\ \bibnamefont
			{Whitney}}, \bibinfo {author} {\bibfnamefont {M.~P.}\ \bibnamefont
			{Levendorf}}, \bibinfo {author} {\bibfnamefont {J.~W.}\ \bibnamefont
			{Kevek}}, \bibinfo {author} {\bibfnamefont {S.}~\bibnamefont {Garg}},
		\bibinfo {author} {\bibfnamefont {J.~S.}\ \bibnamefont {Alden}}, \bibinfo
		{author} {\bibfnamefont {C.~J.}\ \bibnamefont {Hustedt}}, \bibinfo {author}
		{\bibfnamefont {Y.}~\bibnamefont {Zhu}}, \bibinfo {author} {\bibfnamefont
			{J.}~\bibnamefont {Park}}, \bibinfo {author} {\bibfnamefont {P.~L.}\
			\bibnamefont {McEuen}},\ and\ \bibinfo {author} {\bibfnamefont {D.~A.}\
			\bibnamefont {Muller}},\ }\bibfield  {title} {\bibinfo {title} {{Grains and
				grain boundaries in single-layer graphene atomic patchwork quilts}},\ }\href
	{https://doi.org/10.1038/nature09718} {\bibfield  {journal} {\bibinfo
			{journal} {Nature}\ }\textbf {\bibinfo {volume} {469}},\ \bibinfo {pages}
		{389} (\bibinfo {year} {2011})}\BibitemShut {NoStop}%
	\bibitem [{\citenamefont {Lin}\ \emph {et~al.}(2015)\citenamefont {Lin},
		\citenamefont {Komsa}, \citenamefont {Yeh}, \citenamefont {Bj{\"o}rkman},
		\citenamefont {Liang}, \citenamefont {Ho}, \citenamefont {Huang},
		\citenamefont {Chiu}, \citenamefont {Krasheninnikov},\ and\ \citenamefont
		{Suenaga}}]{Lin_2016}%
	\BibitemOpen
	\bibfield  {author} {\bibinfo {author} {\bibfnamefont {Y.-C.}\ \bibnamefont
			{Lin}}, \bibinfo {author} {\bibfnamefont {H.-P.}\ \bibnamefont {Komsa}},
		\bibinfo {author} {\bibfnamefont {C.-H.}\ \bibnamefont {Yeh}}, \bibinfo
		{author} {\bibfnamefont {T.}~\bibnamefont {Bj{\"o}rkman}}, \bibinfo {author}
		{\bibfnamefont {Z.-Y.}\ \bibnamefont {Liang}}, \bibinfo {author}
		{\bibfnamefont {C.-H.}\ \bibnamefont {Ho}}, \bibinfo {author} {\bibfnamefont
			{Y.-S.}\ \bibnamefont {Huang}}, \bibinfo {author} {\bibfnamefont {P.-W.}\
			\bibnamefont {Chiu}}, \bibinfo {author} {\bibfnamefont {A.~V.}\ \bibnamefont
			{Krasheninnikov}},\ and\ \bibinfo {author} {\bibfnamefont {K.}~\bibnamefont
			{Suenaga}},\ }\bibfield  {title} {\bibinfo {title} {{Single-layer ReS$_2$:
				Two-dimensional semiconductor with tunable electronic properties controlled
				by defects}},\ }\href {https://doi.org/10.1021/acsnano.5b04851} {\bibfield
		{journal} {\bibinfo  {journal} {ACS Nano}\ }\textbf {\bibinfo {volume} {9}},\
		\bibinfo {pages} {11249} (\bibinfo {year} {2015})}\BibitemShut {NoStop}%
	\bibitem [{\citenamefont {Kashchiev}(2000)}]{Kashchiev_2000}%
	\BibitemOpen
	\bibfield  {author} {\bibinfo {author} {\bibfnamefont {D.}~\bibnamefont
			{Kashchiev}},\ }\bibfield  {title} {\bibinfo {title} {{Equilibrium}},\ }in\
	\href {https://doi.org/10.1016/B978-075064682-6/50013-5} {\emph {\bibinfo
			{booktitle} {Nucleation}}}\ (\bibinfo  {publisher} {Butterworth-Heinemann},\
	\bibinfo {address} {Oxford},\ \bibinfo {year} {2000})\ Chap.~\bibinfo
	{chapter} {12}, pp.\ \bibinfo {pages} {178--183}\BibitemShut {NoStop}%
	\bibitem [{\citenamefont {Bird}\ \emph {et~al.}(2002)\citenamefont {Bird},
		\citenamefont {Stewart},\ and\ \citenamefont {Lightfoot}}]{Bird_2002}%
	\BibitemOpen
	\bibfield  {author} {\bibinfo {author} {\bibfnamefont {R.~B.}\ \bibnamefont
			{Bird}}, \bibinfo {author} {\bibfnamefont {W.~E.}\ \bibnamefont {Stewart}},\
		and\ \bibinfo {author} {\bibfnamefont {E.~N.}\ \bibnamefont {Lightfoot}},\
	}\href@noop {} {\emph {\bibinfo {title} {{Transport Phenomena}}}},\ \bibinfo
	{edition} {2nd}\ ed.\ (\bibinfo  {publisher} {John Wiley \& Sons},\ \bibinfo
	{address} {New York},\ \bibinfo {year} {2002})\BibitemShut {NoStop}%
	\bibitem [{\citenamefont {Burton}\ \emph {et~al.}(1951)\citenamefont {Burton},
		\citenamefont {Cabrera},\ and\ \citenamefont {Frank}}]{Burton_1951}%
	\BibitemOpen
	\bibfield  {author} {\bibinfo {author} {\bibfnamefont {W.~K.}\ \bibnamefont
			{Burton}}, \bibinfo {author} {\bibfnamefont {N.}~\bibnamefont {Cabrera}},\
		and\ \bibinfo {author} {\bibfnamefont {F.~C.}\ \bibnamefont {Frank}},\
	}\bibfield  {title} {\bibinfo {title} {{The Growth of Crystals and the
				Equilibrium Structure of Their Surfaces}},\ }\href
	{https://doi.org/10.1098/rsta.1951.0006} {\bibfield  {journal} {\bibinfo
			{journal} {Phil. Trans. R. Soc. A}\ }\textbf {\bibinfo {volume} {243}},\
		\bibinfo {pages} {299} (\bibinfo {year} {1951})}\BibitemShut {NoStop}%
	\bibitem [{\citenamefont {Binnewies}\ \emph {et~al.}(2017)\citenamefont
		{Binnewies}, \citenamefont {Schmidt},\ and\ \citenamefont
		{Schmidt}}]{Binnewies_2017}%
	\BibitemOpen
	\bibfield  {author} {\bibinfo {author} {\bibfnamefont {M.}~\bibnamefont
			{Binnewies}}, \bibinfo {author} {\bibfnamefont {M.}~\bibnamefont {Schmidt}},\
		and\ \bibinfo {author} {\bibfnamefont {P.}~\bibnamefont {Schmidt}},\
	}\bibfield  {title} {\bibinfo {title} {{Chemical Vapor Transport Reactions --
				Arguments for Choosing a Suitable Transport Agent}},\ }\href
	{https://doi.org/10.1002/zaac.201700055} {\bibfield  {journal} {\bibinfo
			{journal} {Z. Anorg. Allg. Chem.}\ }\textbf {\bibinfo {volume} {643}},\
		\bibinfo {pages} {1295} (\bibinfo {year} {2017})}\BibitemShut {NoStop}%
	\bibitem [{\citenamefont {Canfield}\ and\ \citenamefont
		{Fisk}(1992)}]{Canfield_1992}%
	\BibitemOpen
	\bibfield  {author} {\bibinfo {author} {\bibfnamefont {P.~C.}\ \bibnamefont
			{Canfield}}\ and\ \bibinfo {author} {\bibfnamefont {Z.}~\bibnamefont
			{Fisk}},\ }\bibfield  {title} {\bibinfo {title} {Growth of single crystals
			from metallic fluxes},\ }\href {https://doi.org/10.1080/13642819208215073}
	{\bibfield  {journal} {\bibinfo  {journal} {Philosophical Magazine B}\
		}\textbf {\bibinfo {volume} {65}},\ \bibinfo {pages} {1117} (\bibinfo {year}
		{1992})}\BibitemShut {NoStop}%
	\bibitem [{\citenamefont {Rudolph}(2003)}]{Rudolph_2003}%
	\BibitemOpen
	\bibfield  {author} {\bibinfo {author} {\bibfnamefont {P.}~\bibnamefont
			{Rudolph}},\ }\href@noop {} {\emph {\bibinfo {title} {{Handbook of Crystal
					Growth}}}},\ \bibinfo {edition} {2nd}\ ed.\ (\bibinfo  {publisher}
	{Elsevier},\ \bibinfo {year} {2003})\BibitemShut {NoStop}%
	\bibitem [{\citenamefont {Lv}\ \emph {et~al.}(2015)\citenamefont {Lv},
		\citenamefont {Robinson}, \citenamefont {Schaak}, \citenamefont {Sun},
		\citenamefont {Sun}, \citenamefont {Mallouk},\ and\ \citenamefont
		{Terrones}}]{Lv_2015}%
	\BibitemOpen
	\bibfield  {author} {\bibinfo {author} {\bibfnamefont {R.}~\bibnamefont
			{Lv}}, \bibinfo {author} {\bibfnamefont {J.~A.}\ \bibnamefont {Robinson}},
		\bibinfo {author} {\bibfnamefont {R.~E.}\ \bibnamefont {Schaak}}, \bibinfo
		{author} {\bibfnamefont {D.}~\bibnamefont {Sun}}, \bibinfo {author}
		{\bibfnamefont {Y.}~\bibnamefont {Sun}}, \bibinfo {author} {\bibfnamefont
			{T.~E.}\ \bibnamefont {Mallouk}},\ and\ \bibinfo {author} {\bibfnamefont
			{M.}~\bibnamefont {Terrones}},\ }\bibfield  {title} {\bibinfo {title}
		{{Transition Metal Dichalcogenides and Beyond: Synthesis, Properties, and
				Applications of Single- and Few-Layer Nanosheets}},\ }\href
	{https://doi.org/10.1021/ar5002846} {\bibfield  {journal} {\bibinfo
			{journal} {Acc. Chem. Res.}\ }\textbf {\bibinfo {volume} {48}},\ \bibinfo
		{pages} {56} (\bibinfo {year} {2015})}\BibitemShut {NoStop}%
	\bibitem [{\citenamefont {Je}\ \emph {et~al.}(2023)\citenamefont {Je},
		\citenamefont {Kim}, \citenamefont {Binh}, \citenamefont {Kim}, \citenamefont
		{Cho}, \citenamefont {Lee}, \citenamefont {Kwon}, \citenamefont {Choi},
		\citenamefont {Lee}, \citenamefont {Nam}, \citenamefont {Lee}, \citenamefont
		{Cho},\ and\ \citenamefont {Park}}]{Je_2023}%
	\BibitemOpen
	\bibfield  {author} {\bibinfo {author} {\bibfnamefont {Y.}~\bibnamefont
			{Je}}, \bibinfo {author} {\bibfnamefont {E.}~\bibnamefont {Kim}}, \bibinfo
		{author} {\bibfnamefont {N.~V.}\ \bibnamefont {Binh}}, \bibinfo {author}
		{\bibfnamefont {H.}~\bibnamefont {Kim}}, \bibinfo {author} {\bibfnamefont
			{S.-y.}\ \bibnamefont {Cho}}, \bibinfo {author} {\bibfnamefont {D.-H.}\
			\bibnamefont {Lee}}, \bibinfo {author} {\bibfnamefont {M.~J.}\ \bibnamefont
			{Kwon}}, \bibinfo {author} {\bibfnamefont {M.}~\bibnamefont {Choi}}, \bibinfo
		{author} {\bibfnamefont {J.~H.}\ \bibnamefont {Lee}}, \bibinfo {author}
		{\bibfnamefont {W.~H.}\ \bibnamefont {Nam}}, \bibinfo {author} {\bibfnamefont
			{Y.}~\bibnamefont {Lee}}, \bibinfo {author} {\bibfnamefont {J.~Y.}\
			\bibnamefont {Cho}},\ and\ \bibinfo {author} {\bibfnamefont {J.~H.}\
			\bibnamefont {Park}},\ }\bibfield  {title} {\bibinfo {title} {{Analysis of
				Physical and Electrical Properties of {NiTe$_2$} Single Crystal Grown via
				Molten Salt Flux Method}},\ }\href
	{https://doi.org/10.1007/s13391-023-00419-2} {\bibfield  {journal} {\bibinfo
			{journal} {Electron. Mater. Lett.}\ }\textbf {\bibinfo {volume} {19}},\
		\bibinfo {pages} {452} (\bibinfo {year} {2023})}\BibitemShut {NoStop}%
	\bibitem [{\citenamefont {Lazarevi\'c}\ \emph {et~al.}(2012)\citenamefont
		{Lazarevi\'c}, \citenamefont {Radonji\'c}, \citenamefont {Hu}, \citenamefont
		{Tanaskovi\'c}, \citenamefont {Petrovic},\ and\ \citenamefont
		{Popovi\'c}}]{Laz_2012}%
	\BibitemOpen
	\bibfield  {author} {\bibinfo {author} {\bibfnamefont {N.}~\bibnamefont
			{Lazarevi\'c}}, \bibinfo {author} {\bibfnamefont {M.~M.}\ \bibnamefont
			{Radonji\'c}}, \bibinfo {author} {\bibfnamefont {R.}~\bibnamefont {Hu}},
		\bibinfo {author} {\bibfnamefont {D.}~\bibnamefont {Tanaskovi\'c}}, \bibinfo
		{author} {\bibfnamefont {C.}~\bibnamefont {Petrovic}},\ and\ \bibinfo
		{author} {\bibfnamefont {Z.~V.}\ \bibnamefont {Popovi\'c}},\ }\bibfield
	{title} {\bibinfo {title} {{Phonon Properties of {CoSb$_2$} Single
				Crystals}},\ }\href {https://doi.org/10.1088/0953-8984/24/13/135402}
	{\bibfield  {journal} {\bibinfo  {journal} {J. Phys.: Condens. Matter}\
		}\textbf {\bibinfo {volume} {24}},\ \bibinfo {pages} {135402} (\bibinfo
		{year} {2012})}\BibitemShut {NoStop}%
	\bibitem [{\citenamefont {Nagao}\ \emph {et~al.}(2021)\citenamefont {Nagao},
		\citenamefont {Miura}, \citenamefont {Maruyama}, \citenamefont {Watauchi},
		\citenamefont {Takano},\ and\ \citenamefont {Tanaka}}]{Nagao_2021}%
	\BibitemOpen
	\bibfield  {author} {\bibinfo {author} {\bibfnamefont {M.}~\bibnamefont
			{Nagao}}, \bibinfo {author} {\bibfnamefont {A.}~\bibnamefont {Miura}},
		\bibinfo {author} {\bibfnamefont {Y.}~\bibnamefont {Maruyama}}, \bibinfo
		{author} {\bibfnamefont {S.}~\bibnamefont {Watauchi}}, \bibinfo {author}
		{\bibfnamefont {Y.}~\bibnamefont {Takano}},\ and\ \bibinfo {author}
		{\bibfnamefont {I.}~\bibnamefont {Tanaka}},\ }\bibfield  {title} {\bibinfo
		{title} {{Cd Additive Effect on Self-Flux Growth of Cs-Intercalated {NbS$_2$}
				Superconducting Single Crystals}},\ }\href
	{https://doi.org/10.1515/znb-2021-0123} {\bibfield  {journal} {\bibinfo
			{journal} {Z. Naturforsch. B}\ }\textbf {\bibinfo {volume} {76}},\ \bibinfo
		{pages} {739} (\bibinfo {year} {2021})}\BibitemShut {NoStop}%
	\bibitem [{\citenamefont {Zhu}\ \emph {et~al.}(2009)\citenamefont {Zhu},
		\citenamefont {Sun}, \citenamefont {Zhang}, \citenamefont {Lei},
		\citenamefont {Li}, \citenamefont {Zhu}, \citenamefont {Yang}, \citenamefont
		{Song},\ and\ \citenamefont {Dai}}]{Zhu_2009}%
	\BibitemOpen
	\bibfield  {author} {\bibinfo {author} {\bibfnamefont {X.~D.}\ \bibnamefont
			{Zhu}}, \bibinfo {author} {\bibfnamefont {Y.~P.}\ \bibnamefont {Sun}},
		\bibinfo {author} {\bibfnamefont {S.~B.}\ \bibnamefont {Zhang}}, \bibinfo
		{author} {\bibfnamefont {H.~C.}\ \bibnamefont {Lei}}, \bibinfo {author}
		{\bibfnamefont {L.~J.}\ \bibnamefont {Li}}, \bibinfo {author} {\bibfnamefont
			{X.~B.}\ \bibnamefont {Zhu}}, \bibinfo {author} {\bibfnamefont {Z.~R.}\
			\bibnamefont {Yang}}, \bibinfo {author} {\bibfnamefont {W.~H.}\ \bibnamefont
			{Song}},\ and\ \bibinfo {author} {\bibfnamefont {J.~M.}\ \bibnamefont
			{Dai}},\ }\bibfield  {title} {\bibinfo {title} {{Superconductivity and Single
				Crystal Growth of Ni$_{0.05}$TaS$_2$}},\ }\href
	{https://doi.org/10.1016/j.ssc.2009.05.007} {\bibfield  {journal} {\bibinfo
			{journal} {Solid State Commun.}\ }\textbf {\bibinfo {volume} {149}},\
		\bibinfo {pages} {1296} (\bibinfo {year} {2009})}\BibitemShut {NoStop}%
	\bibitem [{\citenamefont {Colell}\ \emph {et~al.}(1994)\citenamefont {Colell},
		\citenamefont {Alonso-Vante}, \citenamefont {Fiechter}, \citenamefont
		{Schieck}, \citenamefont {Diesner}, \citenamefont {Henrion},\ and\
		\citenamefont {Tributsch}}]{Colell_1994}%
	\BibitemOpen
	\bibfield  {author} {\bibinfo {author} {\bibfnamefont {H.}~\bibnamefont
			{Colell}}, \bibinfo {author} {\bibfnamefont {N.}~\bibnamefont
			{Alonso-Vante}}, \bibinfo {author} {\bibfnamefont {S.}~\bibnamefont
			{Fiechter}}, \bibinfo {author} {\bibfnamefont {R.}~\bibnamefont {Schieck}},
		\bibinfo {author} {\bibfnamefont {K.}~\bibnamefont {Diesner}}, \bibinfo
		{author} {\bibfnamefont {W.}~\bibnamefont {Henrion}},\ and\ \bibinfo {author}
		{\bibfnamefont {H.}~\bibnamefont {Tributsch}},\ }\bibfield  {title} {\bibinfo
		{title} {{Crystal growth and properties of novel ternary transition metal
				chalcogenide compounds [Ir$_x$Ru$_{1-x}$S$_2$ ($0.005 < x < 0.5$)]}},\ }\href
	{https://doi.org/10.1016/0025-5408(94)90088-4} {\bibfield  {journal}
		{\bibinfo  {journal} {Mater. Res. Bull.}\ }\textbf {\bibinfo {volume} {29}},\
		\bibinfo {pages} {1065} (\bibinfo {year} {1994})}\BibitemShut {NoStop}%
	\bibitem [{\citenamefont {Canfield}\ \emph {et~al.}(2016)\citenamefont
		{Canfield}, \citenamefont {Kong}, \citenamefont {Kaluarachchi},\ and\
		\citenamefont {Jo}}]{Canfield_2016}%
	\BibitemOpen
	\bibfield  {author} {\bibinfo {author} {\bibfnamefont {P.~C.}\ \bibnamefont
			{Canfield}}, \bibinfo {author} {\bibfnamefont {T.}~\bibnamefont {Kong}},
		\bibinfo {author} {\bibfnamefont {U.~S.}\ \bibnamefont {Kaluarachchi}},\ and\
		\bibinfo {author} {\bibfnamefont {N.~H.}\ \bibnamefont {Jo}},\ }\bibfield
	{title} {\bibinfo {title} {{Use of Frit-Disc Crucibles for Routine and
				Exploratory Solution Growth of Single Crystalline Samples}},\ }\href
	{https://doi.org/10.1080/14786435.2015.1122248} {\bibfield  {journal}
		{\bibinfo  {journal} {Philos. Mag.}\ }\textbf {\bibinfo {volume} {96}},\
		\bibinfo {pages} {84} (\bibinfo {year} {2016})}\BibitemShut {NoStop}%
	\bibitem [{\citenamefont {Zhang}\ and\ \citenamefont
		{Wang}(2020)}]{Zhang_2020}%
	\BibitemOpen
	\bibfield  {author} {\bibinfo {author} {\bibfnamefont {Q.}~\bibnamefont
			{Zhang}}\ and\ \bibinfo {author} {\bibfnamefont {X.}~\bibnamefont {Wang}},\
	}\bibfield  {title} {\bibinfo {title} {{Preparation and Crystal Growth of
				Transition Metal Dichalcogenides}},\ }\href
	{https://doi.org/10.1002/zaac.202000111} {\bibfield  {journal} {\bibinfo
			{journal} {Z. Anorg. Allg. Chem.}\ }\textbf {\bibinfo {volume} {646}},\
		\bibinfo {pages} {1047} (\bibinfo {year} {2020})}\BibitemShut {NoStop}%
	\bibitem [{\citenamefont {Chareev}\ \emph {et~al.}(2020)\citenamefont
		{Chareev}, \citenamefont {Evstigneeva}, \citenamefont {Phuyal}, \citenamefont
		{Man}, \citenamefont {Rensmo}, \citenamefont {Vasiliev},\ and\ \citenamefont
		{Abdel-Hafiez}}]{Chareev_2020}%
	\BibitemOpen
	\bibfield  {author} {\bibinfo {author} {\bibfnamefont {D.~A.}\ \bibnamefont
			{Chareev}}, \bibinfo {author} {\bibfnamefont {P.~V.}\ \bibnamefont
			{Evstigneeva}}, \bibinfo {author} {\bibfnamefont {D.}~\bibnamefont {Phuyal}},
		\bibinfo {author} {\bibfnamefont {G.}~\bibnamefont {Man}}, \bibinfo {author}
		{\bibfnamefont {H.}~\bibnamefont {Rensmo}}, \bibinfo {author} {\bibfnamefont
			{A.~N.}\ \bibnamefont {Vasiliev}},\ and\ \bibinfo {author} {\bibfnamefont
			{M.}~\bibnamefont {Abdel-Hafiez}},\ }\bibfield  {title} {\bibinfo {title}
		{{Growth of Transition-Metal Dichalcogenides by Solvent Evaporation
				Technique}},\ }\href {https://doi.org/10.1021/acs.cgd.0c00980} {\bibfield
		{journal} {\bibinfo  {journal} {Cryst. Growth Des.}\ }\textbf {\bibinfo
			{volume} {20}},\ \bibinfo {pages} {8280} (\bibinfo {year}
		{2020})}\BibitemShut {NoStop}%
	\bibitem [{\citenamefont {Venables}\ \emph {et~al.}(1984)\citenamefont
		{Venables}, \citenamefont {Spiller},\ and\ \citenamefont
		{Hanb\"ucken}}]{Venables_1984}%
	\BibitemOpen
	\bibfield  {author} {\bibinfo {author} {\bibfnamefont {J.~A.}\ \bibnamefont
			{Venables}}, \bibinfo {author} {\bibfnamefont {G.~D.~T.}\ \bibnamefont
			{Spiller}},\ and\ \bibinfo {author} {\bibfnamefont {M.}~\bibnamefont
			{Hanb\"ucken}},\ }\bibfield  {title} {\bibinfo {title} {{Nucleation and
				growth of thin films}},\ }\href {https://doi.org/10.1088/0034-4885/47/4/002}
	{\bibfield  {journal} {\bibinfo  {journal} {Reports on Progress in Physics}\
		}\textbf {\bibinfo {volume} {47}},\ \bibinfo {pages} {399} (\bibinfo {year}
		{1984})}\BibitemShut {NoStop}%
	\bibitem [{\citenamefont {Raja}\ \emph {et~al.}(2017)\citenamefont {Raja},
		\citenamefont {Chaves}, \citenamefont {Yu}, \citenamefont {Arefe},
		\citenamefont {Hill}, \citenamefont {Rigosi}, \citenamefont {Berkelbach},
		\citenamefont {Nagler}, \citenamefont {Sch{\"u}ller}, \citenamefont {Korn},
		\citenamefont {Nuckolls}, \citenamefont {Hone}, \citenamefont {Brus},
		\citenamefont {Heinz}, \citenamefont {Reichman},\ and\ \citenamefont
		{Chernikov}}]{Raja_2017}%
	\BibitemOpen
	\bibfield  {author} {\bibinfo {author} {\bibfnamefont {A.}~\bibnamefont
			{Raja}}, \bibinfo {author} {\bibfnamefont {A.}~\bibnamefont {Chaves}},
		\bibinfo {author} {\bibfnamefont {J.}~\bibnamefont {Yu}}, \bibinfo {author}
		{\bibfnamefont {G.}~\bibnamefont {Arefe}}, \bibinfo {author} {\bibfnamefont
			{H.~M.}\ \bibnamefont {Hill}}, \bibinfo {author} {\bibfnamefont {A.~F.}\
			\bibnamefont {Rigosi}}, \bibinfo {author} {\bibfnamefont {T.~C.}\
			\bibnamefont {Berkelbach}}, \bibinfo {author} {\bibfnamefont
			{P.}~\bibnamefont {Nagler}}, \bibinfo {author} {\bibfnamefont
			{C.}~\bibnamefont {Sch{\"u}ller}}, \bibinfo {author} {\bibfnamefont
			{T.}~\bibnamefont {Korn}}, \bibinfo {author} {\bibfnamefont {C.}~\bibnamefont
			{Nuckolls}}, \bibinfo {author} {\bibfnamefont {J.}~\bibnamefont {Hone}},
		\bibinfo {author} {\bibfnamefont {L.~E.}\ \bibnamefont {Brus}}, \bibinfo
		{author} {\bibfnamefont {T.~F.}\ \bibnamefont {Heinz}}, \bibinfo {author}
		{\bibfnamefont {D.~R.}\ \bibnamefont {Reichman}},\ and\ \bibinfo {author}
		{\bibfnamefont {A.}~\bibnamefont {Chernikov}},\ }\bibfield  {title} {\bibinfo
		{title} {Coulomb engineering of the bandgap and excitons in two-dimensional
			materials},\ }\href {https://doi.org/10.1038/ncomms15251} {\bibfield
		{journal} {\bibinfo  {journal} {Nature Communications}\ }\textbf {\bibinfo
			{volume} {8}},\ \bibinfo {pages} {15251} (\bibinfo {year}
		{2017})}\BibitemShut {NoStop}%
	\bibitem [{\citenamefont {van~der Zande}\ \emph {et~al.}(2013)\citenamefont
		{van~der Zande}, \citenamefont {Huang}, \citenamefont {Chenet}, \citenamefont
		{Berkelbach}, \citenamefont {You}, \citenamefont {Lee}, \citenamefont
		{Heinz}, \citenamefont {Reichman}, \citenamefont {Muller},\ and\
		\citenamefont {Hone}}]{VanDerZande_2013}%
	\BibitemOpen
	\bibfield  {author} {\bibinfo {author} {\bibfnamefont {A.~M.}\ \bibnamefont
			{van~der Zande}}, \bibinfo {author} {\bibfnamefont {P.~Y.}\ \bibnamefont
			{Huang}}, \bibinfo {author} {\bibfnamefont {D.~A.}\ \bibnamefont {Chenet}},
		\bibinfo {author} {\bibfnamefont {T.~C.}\ \bibnamefont {Berkelbach}},
		\bibinfo {author} {\bibfnamefont {Y.}~\bibnamefont {You}}, \bibinfo {author}
		{\bibfnamefont {G.-H.}\ \bibnamefont {Lee}}, \bibinfo {author} {\bibfnamefont
			{T.~F.}\ \bibnamefont {Heinz}}, \bibinfo {author} {\bibfnamefont {D.~R.}\
			\bibnamefont {Reichman}}, \bibinfo {author} {\bibfnamefont {D.~A.}\
			\bibnamefont {Muller}},\ and\ \bibinfo {author} {\bibfnamefont {J.~C.}\
			\bibnamefont {Hone}},\ }\bibfield  {title} {\bibinfo {title} {{Grains and
				Grain Boundaries in Highly Crystalline Monolayer Molybdenum Disulphide}},\
	}\href {https://doi.org/10.1038/nmat3633} {\bibfield  {journal} {\bibinfo
			{journal} {Nat. Mater.}\ }\textbf {\bibinfo {volume} {12}},\ \bibinfo {pages}
		{554} (\bibinfo {year} {2013})}\BibitemShut {NoStop}%
	\bibitem [{\citenamefont {Xue}\ \emph {et~al.}(2024)\citenamefont {Xue},
		\citenamefont {Qin}, \citenamefont {Ma}, \citenamefont {Yin}, \citenamefont
		{Liu},\ and\ \citenamefont {Liu}}]{Xue_2024}%
	\BibitemOpen
	\bibfield  {author} {\bibinfo {author} {\bibfnamefont {G.}~\bibnamefont
			{Xue}}, \bibinfo {author} {\bibfnamefont {B.}~\bibnamefont {Qin}}, \bibinfo
		{author} {\bibfnamefont {C.}~\bibnamefont {Ma}}, \bibinfo {author}
		{\bibfnamefont {P.}~\bibnamefont {Yin}}, \bibinfo {author} {\bibfnamefont
			{C.}~\bibnamefont {Liu}},\ and\ \bibinfo {author} {\bibfnamefont
			{K.}~\bibnamefont {Liu}},\ }\bibfield  {title} {\bibinfo {title} {{Large-Area
				Epitaxial Growth of Transition Metal Dichalcogenides}},\ }\href
	{https://doi.org/10.1021/acs.chemrev.3c00851} {\bibfield  {journal} {\bibinfo
			{journal} {Chem. Rev.}\ }\textbf {\bibinfo {volume} {124}},\ \bibinfo
		{pages} {9785} (\bibinfo {year} {2024})}\BibitemShut {NoStop}%
	\bibitem [{\citenamefont {Cheng}\ \emph {et~al.}(2020)\citenamefont {Cheng},
		\citenamefont {Zhang}, \citenamefont {Zhang}, \citenamefont {Gou},
		\citenamefont {Wong},\ and\ \citenamefont {Chen}}]{Cheng_2020}%
	\BibitemOpen
	\bibfield  {author} {\bibinfo {author} {\bibfnamefont {P.}~\bibnamefont
			{Cheng}}, \bibinfo {author} {\bibfnamefont {W.}~\bibnamefont {Zhang}},
		\bibinfo {author} {\bibfnamefont {L.}~\bibnamefont {Zhang}}, \bibinfo
		{author} {\bibfnamefont {J.}~\bibnamefont {Gou}}, \bibinfo {author}
		{\bibfnamefont {P.~K.~J.}\ \bibnamefont {Wong}},\ and\ \bibinfo {author}
		{\bibfnamefont {L.}~\bibnamefont {Chen}},\ }\bibfield  {title} {\bibinfo
		{title} {{Molecular Beam Epitaxy Fabrication of Two-Dimensional Materials}},\
	}in\ \href {https://doi.org/10.1016/B978-0-12-816187-6.00004-2} {\emph
		{\bibinfo {booktitle} {2D Semiconductor Materials and Devices}}},\ \bibinfo
	{series and number} {Materials Today},\ \bibinfo {editor} {edited by\
		\bibinfo {editor} {\bibfnamefont {D.}~\bibnamefont {Chi}}, \bibinfo {editor}
		{\bibfnamefont {K.~E.~J.}\ \bibnamefont {Goh}},\ and\ \bibinfo {editor}
		{\bibfnamefont {A.~T.~S.}\ \bibnamefont {Wee}}}\ (\bibinfo  {publisher}
	{Elsevier},\ \bibinfo {year} {2020})\ pp.\ \bibinfo {pages}
	{103--134}\BibitemShut {NoStop}%
	\bibitem [{\citenamefont {Vishwanath}\ \emph {et~al.}(2018)\citenamefont
		{Vishwanath}, \citenamefont {Dang},\ and\ \citenamefont
		{Xing}}]{Suresh_2018}%
	\BibitemOpen
	\bibfield  {author} {\bibinfo {author} {\bibfnamefont {S.}~\bibnamefont
			{Vishwanath}}, \bibinfo {author} {\bibfnamefont {P.}~\bibnamefont {Dang}},\
		and\ \bibinfo {author} {\bibfnamefont {H.~G.}\ \bibnamefont {Xing}},\
	}\bibfield  {title} {\bibinfo {title} {{Challenges and Opportunities in
				Molecular Beam Epitaxy Growth of 2D Crystals: An Overview}},\ }in\ \href
	{https://doi.org/10.1016/B978-0-12-812136-8.00017-7} {\emph {\bibinfo
			{booktitle} {Molecular Beam Epitaxy (Second Edition)}}},\ \bibinfo {editor}
	{edited by\ \bibinfo {editor} {\bibfnamefont {M.}~\bibnamefont {Henini}}}\
	(\bibinfo  {publisher} {Elsevier},\ \bibinfo {year} {2018})\ \bibinfo
	{edition} {second edition}\ ed.,\ pp.\ \bibinfo {pages}
	{443--485}\BibitemShut {NoStop}%
	\bibitem [{\citenamefont {Singh}\ \emph {et~al.}(2020)\citenamefont {Singh},
		\citenamefont {Nanda},\ and\ \citenamefont {Krupanidhi}}]{Singh_2020}%
	\BibitemOpen
	\bibfield  {author} {\bibinfo {author} {\bibfnamefont {D.~K.}\ \bibnamefont
			{Singh}}, \bibinfo {author} {\bibfnamefont {K.~K.}\ \bibnamefont {Nanda}},\
		and\ \bibinfo {author} {\bibfnamefont {S.~B.}\ \bibnamefont {Krupanidhi}},\
	}\bibfield  {title} {\bibinfo {title} {{Pulsed Laser Deposition of Transition
				Metal Dichalcogenides-Based Heterostructures for Efficient Photodetection}},\
	}in\ \href {https://doi.org/10.5772/intechopen.94236} {\emph {\bibinfo
			{booktitle} {Practical Applications of Laser Ablation}}},\ \bibinfo {editor}
	{edited by\ \bibinfo {editor} {\bibfnamefont {D.}~\bibnamefont {Yang}}}\
	(\bibinfo  {publisher} {IntechOpen},\ \bibinfo {address} {London},\ \bibinfo
	{year} {2020})\ Chap.~\bibinfo {chapter} {2}\BibitemShut {NoStop}%
	\bibitem [{\citenamefont {Rai}\ \emph {et~al.}(2020)\citenamefont {Rai},
		\citenamefont {Pérez-Pacheco}, \citenamefont {Quispe-Siccha}, \citenamefont
		{Glavin},\ and\ \citenamefont {Muratore}}]{Rai_2020}%
	\BibitemOpen
	\bibfield  {author} {\bibinfo {author} {\bibfnamefont {R.~H.}\ \bibnamefont
			{Rai}}, \bibinfo {author} {\bibfnamefont {A.}~\bibnamefont {Pérez-Pacheco}},
		\bibinfo {author} {\bibfnamefont {R.}~\bibnamefont {Quispe-Siccha}}, \bibinfo
		{author} {\bibfnamefont {N.~R.}\ \bibnamefont {Glavin}},\ and\ \bibinfo
		{author} {\bibfnamefont {C.}~\bibnamefont {Muratore}},\ }\bibfield  {title}
	{\bibinfo {title} {{Pulsed Laser Annealing of Amorphous Two-Dimensional
				Transition Metal Dichalcogenides}},\ }\href
	{https://doi.org/10.1116/6.0000253} {\bibfield  {journal} {\bibinfo
			{journal} {J. Vac. Sci. Technol. A}\ }\textbf {\bibinfo {volume} {38}},\
		\bibinfo {pages} {052201} (\bibinfo {year} {2020})}\BibitemShut {NoStop}%
	\bibitem [{\citenamefont {Wong}\ \emph {et~al.}(2017)\citenamefont {Wong},
		\citenamefont {Ng}, \citenamefont {Wong}, \citenamefont {Cheng},
		\citenamefont {Mak},\ and\ \citenamefont {Leung}}]{Wong_2017}%
	\BibitemOpen
	\bibfield  {author} {\bibinfo {author} {\bibfnamefont {W.~C.}\ \bibnamefont
			{Wong}}, \bibinfo {author} {\bibfnamefont {S.~M.}\ \bibnamefont {Ng}},
		\bibinfo {author} {\bibfnamefont {H.~F.}\ \bibnamefont {Wong}}, \bibinfo
		{author} {\bibfnamefont {W.~F.}\ \bibnamefont {Cheng}}, \bibinfo {author}
		{\bibfnamefont {C.~L.}\ \bibnamefont {Mak}},\ and\ \bibinfo {author}
		{\bibfnamefont {C.~W.}\ \bibnamefont {Leung}},\ }\bibfield  {title} {\bibinfo
		{title} {{Effect of Post-Annealing on Sputtered {MoS$_2$} Films}},\ }\href
	{https://doi.org/10.1016/j.sse.2017.07.009} {\bibfield  {journal} {\bibinfo
			{journal} {Solid-State Electron.}\ }\textbf {\bibinfo {volume} {138}},\
		\bibinfo {pages} {62} (\bibinfo {year} {2017})},\ \bibinfo {note} {selected
		papers from INEC 2016 and ISNE 2016}\BibitemShut {NoStop}%
	\bibitem [{\citenamefont {Hussain}\ \emph {et~al.}(2016)\citenamefont
		{Hussain}, \citenamefont {Shehzad}, \citenamefont {Vikraman}, \citenamefont
		{Khan}, \citenamefont {Singh}, \citenamefont {Choi}, \citenamefont {Seo},
		\citenamefont {Eom}, \citenamefont {Lee},\ and\ \citenamefont
		{Jung}}]{Husaain_2016}%
	\BibitemOpen
	\bibfield  {author} {\bibinfo {author} {\bibfnamefont {S.}~\bibnamefont
			{Hussain}}, \bibinfo {author} {\bibfnamefont {M.~A.}\ \bibnamefont
			{Shehzad}}, \bibinfo {author} {\bibfnamefont {D.}~\bibnamefont {Vikraman}},
		\bibinfo {author} {\bibfnamefont {M.~F.}\ \bibnamefont {Khan}}, \bibinfo
		{author} {\bibfnamefont {J.}~\bibnamefont {Singh}}, \bibinfo {author}
		{\bibfnamefont {D.-C.}\ \bibnamefont {Choi}}, \bibinfo {author}
		{\bibfnamefont {Y.}~\bibnamefont {Seo}}, \bibinfo {author} {\bibfnamefont
			{J.}~\bibnamefont {Eom}}, \bibinfo {author} {\bibfnamefont {W.-G.}\
			\bibnamefont {Lee}},\ and\ \bibinfo {author} {\bibfnamefont {J.}~\bibnamefont
			{Jung}},\ }\bibfield  {title} {\bibinfo {title} {{Synthesis and
				Characterization of Large-Area and Continuous {MoS$_2$} Atomic Layers by RF
				Magnetron Sputtering}},\ }\href {https://doi.org/10.1039/C5NR09032F}
	{\bibfield  {journal} {\bibinfo  {journal} {Nanoscale}\ }\textbf {\bibinfo
			{volume} {8}},\ \bibinfo {pages} {4340} (\bibinfo {year} {2016})}\BibitemShut
	{NoStop}%
	\bibitem [{\citenamefont {Zhao}\ \emph {et~al.}(2019)\citenamefont {Zhao},
		\citenamefont {Lin}, \citenamefont {Xiao}, \citenamefont {Huang},
		\citenamefont {Yao}, \citenamefont {Yan}, \citenamefont {Xing}, \citenamefont
		{Zhang}, \citenamefont {Li}, \citenamefont {Hoshino}, \citenamefont {Wang},
		\citenamefont {Zhou}, \citenamefont {Gu}, \citenamefont {Bahramy},
		\citenamefont {Yao}, \citenamefont {Nagaosa}, \citenamefont {Xue},
		\citenamefont {Law}, \citenamefont {Chen},\ and\ \citenamefont
		{Ji}}]{Zhao_2019}%
	\BibitemOpen
	\bibfield  {author} {\bibinfo {author} {\bibfnamefont {K.}~\bibnamefont
			{Zhao}}, \bibinfo {author} {\bibfnamefont {H.}~\bibnamefont {Lin}}, \bibinfo
		{author} {\bibfnamefont {X.}~\bibnamefont {Xiao}}, \bibinfo {author}
		{\bibfnamefont {W.}~\bibnamefont {Huang}}, \bibinfo {author} {\bibfnamefont
			{W.}~\bibnamefont {Yao}}, \bibinfo {author} {\bibfnamefont {M.}~\bibnamefont
			{Yan}}, \bibinfo {author} {\bibfnamefont {Y.}~\bibnamefont {Xing}}, \bibinfo
		{author} {\bibfnamefont {Q.}~\bibnamefont {Zhang}}, \bibinfo {author}
		{\bibfnamefont {Z.-X.}\ \bibnamefont {Li}}, \bibinfo {author} {\bibfnamefont
			{S.}~\bibnamefont {Hoshino}}, \bibinfo {author} {\bibfnamefont
			{J.}~\bibnamefont {Wang}}, \bibinfo {author} {\bibfnamefont {S.}~\bibnamefont
			{Zhou}}, \bibinfo {author} {\bibfnamefont {L.}~\bibnamefont {Gu}}, \bibinfo
		{author} {\bibfnamefont {M.~S.}\ \bibnamefont {Bahramy}}, \bibinfo {author}
		{\bibfnamefont {H.}~\bibnamefont {Yao}}, \bibinfo {author} {\bibfnamefont
			{N.}~\bibnamefont {Nagaosa}}, \bibinfo {author} {\bibfnamefont {Q.-K.}\
			\bibnamefont {Xue}}, \bibinfo {author} {\bibfnamefont {K.~T.}\ \bibnamefont
			{Law}}, \bibinfo {author} {\bibfnamefont {X.}~\bibnamefont {Chen}},\ and\
		\bibinfo {author} {\bibfnamefont {S.-H.}\ \bibnamefont {Ji}},\ }\bibfield
	{title} {\bibinfo {title} {{Disorder-induced multifractal superconductivity
				in monolayer niobium dichalcogenides}},\ }\href
	{https://doi.org/10.1038/s41567-019-0570-0} {\bibfield  {journal} {\bibinfo
			{journal} {Nature Physics}\ }\textbf {\bibinfo {volume} {15}},\ \bibinfo
		{pages} {904} (\bibinfo {year} {2019})}\BibitemShut {NoStop}%
	\bibitem [{\citenamefont {Choe}\ \emph {et~al.}(2016)\citenamefont {Choe},
		\citenamefont {Sung},\ and\ \citenamefont {Chang}}]{Choe_2016}%
	\BibitemOpen
	\bibfield  {author} {\bibinfo {author} {\bibfnamefont {D.-H.}\ \bibnamefont
			{Choe}}, \bibinfo {author} {\bibfnamefont {H.-J.}\ \bibnamefont {Sung}},\
		and\ \bibinfo {author} {\bibfnamefont {K.~J.}\ \bibnamefont {Chang}},\
	}\bibfield  {title} {\bibinfo {title} {{Understanding topological phase
				transition in monolayer transition metal dichalcogenides}},\ }\href
	{https://doi.org/10.1103/PhysRevB.93.125109} {\bibfield  {journal} {\bibinfo
			{journal} {Phys. Rev. B}\ }\textbf {\bibinfo {volume} {93}},\ \bibinfo
		{pages} {125109} (\bibinfo {year} {2016})}\BibitemShut {NoStop}%
	\bibitem [{\citenamefont {Jaques}\ \emph {et~al.}(2026)\citenamefont {Jaques},
		\citenamefont {Woellner}, \citenamefont {Sassi}, \citenamefont
		{Pereira~Junior}, \citenamefont {Ribeiro~Junior}, \citenamefont {Ajayan},\
		and\ \citenamefont {Galv{\~a}o}}]{Jaques_2026}%
	\BibitemOpen
	\bibfield  {author} {\bibinfo {author} {\bibfnamefont {Y.~M.}\ \bibnamefont
			{Jaques}}, \bibinfo {author} {\bibfnamefont {C.~F.}\ \bibnamefont
			{Woellner}}, \bibinfo {author} {\bibfnamefont {L.~M.}\ \bibnamefont {Sassi}},
		\bibinfo {author} {\bibfnamefont {M.~L.}\ \bibnamefont {Pereira~Junior}},
		\bibinfo {author} {\bibfnamefont {L.~A.}\ \bibnamefont {Ribeiro~Junior}},
		\bibinfo {author} {\bibfnamefont {P.~M.}\ \bibnamefont {Ajayan}},\ and\
		\bibinfo {author} {\bibfnamefont {D.~S.}\ \bibnamefont {Galv{\~a}o}},\
	}\bibfield  {title} {\bibinfo {title} {{Structural stability of
				sulfur-depleted MoS$_2$}},\ }\bibfield  {journal} {\bibinfo  {journal} {ACS
			Nanoscience Au}\ }\href {https://doi.org/10.1021/acsnanoscienceau.5c00172}
	{10.1021/acsnanoscienceau.5c00172} (\bibinfo {year} {2026})\BibitemShut
	{NoStop}%
	\bibitem [{\citenamefont {Cho}\ \emph {et~al.}(2018)\citenamefont {Cho},
		\citenamefont {Ko{\'n}czykowski}, \citenamefont {Teknowijoyo}, \citenamefont
		{Tanatar}, \citenamefont {Guss}, \citenamefont {Gartin}, \citenamefont
		{Wilde}, \citenamefont {Kreyssig}, \citenamefont {McQueeney}, \citenamefont
		{Goldman}, \citenamefont {Mishra}, \citenamefont {Hirschfeld},\ and\
		\citenamefont {Prozorov}}]{Cho_2018}%
	\BibitemOpen
	\bibfield  {author} {\bibinfo {author} {\bibfnamefont {K.}~\bibnamefont
			{Cho}}, \bibinfo {author} {\bibfnamefont {M.}~\bibnamefont
			{Ko{\'n}czykowski}}, \bibinfo {author} {\bibfnamefont {S.}~\bibnamefont
			{Teknowijoyo}}, \bibinfo {author} {\bibfnamefont {M.~A.}\ \bibnamefont
			{Tanatar}}, \bibinfo {author} {\bibfnamefont {J.}~\bibnamefont {Guss}},
		\bibinfo {author} {\bibfnamefont {P.~B.}\ \bibnamefont {Gartin}}, \bibinfo
		{author} {\bibfnamefont {J.~M.}\ \bibnamefont {Wilde}}, \bibinfo {author}
		{\bibfnamefont {A.}~\bibnamefont {Kreyssig}}, \bibinfo {author}
		{\bibfnamefont {R.~J.}\ \bibnamefont {McQueeney}}, \bibinfo {author}
		{\bibfnamefont {A.~I.}\ \bibnamefont {Goldman}}, \bibinfo {author}
		{\bibfnamefont {V.}~\bibnamefont {Mishra}}, \bibinfo {author} {\bibfnamefont
			{P.~J.}\ \bibnamefont {Hirschfeld}},\ and\ \bibinfo {author} {\bibfnamefont
			{R.}~\bibnamefont {Prozorov}},\ }\bibfield  {title} {\bibinfo {title} {{Using
				controlled disorder to probe the interplay between charge order and
				superconductivity in NbSe$_2$}},\ }\href
	{https://doi.org/10.1038/s41467-018-05153-0} {\bibfield  {journal} {\bibinfo
			{journal} {Nature Communications}\ }\textbf {\bibinfo {volume} {9}},\
		\bibinfo {pages} {2796} (\bibinfo {year} {2018})}\BibitemShut {NoStop}%
	\bibitem [{\citenamefont {Saini}\ \emph {et~al.}(2023)\citenamefont {Saini},
		\citenamefont {Karn}, \citenamefont {Kumar}, \citenamefont {Aloysius},\ and\
		\citenamefont {Awana}}]{Saini_2023}%
	\BibitemOpen
	\bibfield  {author} {\bibinfo {author} {\bibfnamefont {A.}~\bibnamefont
			{Saini}}, \bibinfo {author} {\bibfnamefont {N.~K.}\ \bibnamefont {Karn}},
		\bibinfo {author} {\bibfnamefont {K.}~\bibnamefont {Kumar}}, \bibinfo
		{author} {\bibfnamefont {R.~P.}\ \bibnamefont {Aloysius}},\ and\ \bibinfo
		{author} {\bibfnamefont {V.~P.~S.}\ \bibnamefont {Awana}},\ }\bibfield
	{title} {\bibinfo {title} {{Drastic suppression of CDW (charge density wave)
				by Pd addition in TiSe$_2$}},\ }\href
	{https://doi.org/10.1007/s10948-022-06453-9} {\bibfield  {journal} {\bibinfo
			{journal} {Journal of Superconductivity and Novel Magnetism}\ }\textbf
		{\bibinfo {volume} {36}},\ \bibinfo {pages} {3} (\bibinfo {year}
		{2023})}\BibitemShut {NoStop}%
	\bibitem [{\citenamefont {Paul}\ \emph {et~al.}(2023)\citenamefont {Paul},
		\citenamefont {Talukdar}, \citenamefont {Singh},\ and\ \citenamefont
		{Saha}}]{Paul_2023}%
	\BibitemOpen
	\bibfield  {author} {\bibinfo {author} {\bibfnamefont {S.}~\bibnamefont
			{Paul}}, \bibinfo {author} {\bibfnamefont {S.}~\bibnamefont {Talukdar}},
		\bibinfo {author} {\bibfnamefont {R.~S.}\ \bibnamefont {Singh}},\ and\
		\bibinfo {author} {\bibfnamefont {S.}~\bibnamefont {Saha}},\ }\bibfield
	{title} {\bibinfo {title} {{Topological phase transition in MoTe$_2$: A
				review}},\ }\href {https://doi.org/10.1002/pssr.202200420} {\bibfield
		{journal} {\bibinfo  {journal} {Physica Status Solidi (RRL) -- Rapid Research
				Letters}\ }\textbf {\bibinfo {volume} {17}},\ \bibinfo {pages} {2200420}
		(\bibinfo {year} {2023})}\BibitemShut {NoStop}%
	\bibitem [{\citenamefont {Sokolikova}\ and\ \citenamefont
		{Mattevi}(2020)}]{Sokolikova_2020}%
	\BibitemOpen
	\bibfield  {author} {\bibinfo {author} {\bibfnamefont {M.~S.}\ \bibnamefont
			{Sokolikova}}\ and\ \bibinfo {author} {\bibfnamefont {C.}~\bibnamefont
			{Mattevi}},\ }\bibfield  {title} {\bibinfo {title} {{Direct synthesis of
				metastable phases of 2D transition metal dichalcogenides}},\ }\href
	{https://doi.org/10.1039/D0CS00143K} {\bibfield  {journal} {\bibinfo
			{journal} {Chemical Society Reviews}\ }\textbf {\bibinfo {volume} {49}},\
		\bibinfo {pages} {3952} (\bibinfo {year} {2020})}\BibitemShut {NoStop}%
	\bibitem [{\citenamefont {Yu}\ \emph {et~al.}(2011)\citenamefont {Yu},
		\citenamefont {Jauregui}, \citenamefont {Wu}, \citenamefont {Colby},
		\citenamefont {Tian}, \citenamefont {Su}, \citenamefont {Cao}, \citenamefont
		{Liu}, \citenamefont {Pandey}, \citenamefont {Wei}, \citenamefont {Chung},
		\citenamefont {Peng}, \citenamefont {Guisinger}, \citenamefont {Stach},
		\citenamefont {Bao}, \citenamefont {Pei},\ and\ \citenamefont
		{Chen}}]{Yu_2011}%
	\BibitemOpen
	\bibfield  {author} {\bibinfo {author} {\bibfnamefont {Q.}~\bibnamefont
			{Yu}}, \bibinfo {author} {\bibfnamefont {L.~A.}\ \bibnamefont {Jauregui}},
		\bibinfo {author} {\bibfnamefont {W.}~\bibnamefont {Wu}}, \bibinfo {author}
		{\bibfnamefont {R.}~\bibnamefont {Colby}}, \bibinfo {author} {\bibfnamefont
			{J.}~\bibnamefont {Tian}}, \bibinfo {author} {\bibfnamefont {Z.}~\bibnamefont
			{Su}}, \bibinfo {author} {\bibfnamefont {H.}~\bibnamefont {Cao}}, \bibinfo
		{author} {\bibfnamefont {Z.}~\bibnamefont {Liu}}, \bibinfo {author}
		{\bibfnamefont {D.}~\bibnamefont {Pandey}}, \bibinfo {author} {\bibfnamefont
			{D.}~\bibnamefont {Wei}}, \bibinfo {author} {\bibfnamefont {T.~F.}\
			\bibnamefont {Chung}}, \bibinfo {author} {\bibfnamefont {P.}~\bibnamefont
			{Peng}}, \bibinfo {author} {\bibfnamefont {N.~P.}\ \bibnamefont {Guisinger}},
		\bibinfo {author} {\bibfnamefont {E.~A.}\ \bibnamefont {Stach}}, \bibinfo
		{author} {\bibfnamefont {J.}~\bibnamefont {Bao}}, \bibinfo {author}
		{\bibfnamefont {S.-S.}\ \bibnamefont {Pei}},\ and\ \bibinfo {author}
		{\bibfnamefont {Y.~P.}\ \bibnamefont {Chen}},\ }\bibfield  {title} {\bibinfo
		{title} {{Control and characterization of individual grains and grain
				boundaries in graphene grown by chemical vapour deposition}},\ }\href
	{https://doi.org/10.1038/nmat3010} {\bibfield  {journal} {\bibinfo  {journal}
			{Nature Materials}\ }\textbf {\bibinfo {volume} {10}},\ \bibinfo {pages}
		{443} (\bibinfo {year} {2011})}\BibitemShut {NoStop}%
	\bibitem [{\citenamefont {Ahn}\ \emph {et~al.}(2017)\citenamefont {Ahn},
		\citenamefont {Amani}, \citenamefont {Rasool}, \citenamefont {Lien},
		\citenamefont {Mastandrea}, \citenamefont {Ager~III}, \citenamefont {Dubey},
		\citenamefont {Chrzan}, \citenamefont {Minor},\ and\ \citenamefont
		{Javey}}]{Ahn_2017}%
	\BibitemOpen
	\bibfield  {author} {\bibinfo {author} {\bibfnamefont {G.~H.}\ \bibnamefont
			{Ahn}}, \bibinfo {author} {\bibfnamefont {M.}~\bibnamefont {Amani}}, \bibinfo
		{author} {\bibfnamefont {H.}~\bibnamefont {Rasool}}, \bibinfo {author}
		{\bibfnamefont {D.-H.}\ \bibnamefont {Lien}}, \bibinfo {author}
		{\bibfnamefont {J.~P.}\ \bibnamefont {Mastandrea}}, \bibinfo {author}
		{\bibfnamefont {J.~W.}\ \bibnamefont {Ager~III}}, \bibinfo {author}
		{\bibfnamefont {M.}~\bibnamefont {Dubey}}, \bibinfo {author} {\bibfnamefont
			{D.~C.}\ \bibnamefont {Chrzan}}, \bibinfo {author} {\bibfnamefont {A.~M.}\
			\bibnamefont {Minor}},\ and\ \bibinfo {author} {\bibfnamefont
			{A.}~\bibnamefont {Javey}},\ }\bibfield  {title} {\bibinfo {title}
		{{Strain-engineered growth of two-dimensional materials}},\ }\href
	{https://doi.org/10.1038/s41467-017-00516-5} {\bibfield  {journal} {\bibinfo
			{journal} {Nature Communications}\ }\textbf {\bibinfo {volume} {8}},\
		\bibinfo {pages} {608} (\bibinfo {year} {2017})}\BibitemShut {NoStop}%
	\bibitem [{\citenamefont {Zunger}\ and\ \citenamefont
		{Malyi}(2021)}]{Zunger_2021}%
	\BibitemOpen
	\bibfield  {author} {\bibinfo {author} {\bibfnamefont {A.}~\bibnamefont
			{Zunger}}\ and\ \bibinfo {author} {\bibfnamefont {O.~I.}\ \bibnamefont
			{Malyi}},\ }\bibfield  {title} {\bibinfo {title} {{Understanding doping of
				quantum materials}},\ }\href {https://doi.org/10.1021/acs.chemrev.0c00608}
	{\bibfield  {journal} {\bibinfo  {journal} {Chemical Reviews}\ }\textbf
		{\bibinfo {volume} {121}},\ \bibinfo {pages} {3031} (\bibinfo {year}
		{2021})}\BibitemShut {NoStop}%
	\bibitem [{\citenamefont {Shuang}\ \emph {et~al.}(2026)\citenamefont {Shuang},
		\citenamefont {Ocampo}, \citenamefont {Namakian}, \citenamefont {Ghasemi},
		\citenamefont {Dey},\ and\ \citenamefont {Gao}}]{Shuang_2026}%
	\BibitemOpen
	\bibfield  {author} {\bibinfo {author} {\bibfnamefont {F.}~\bibnamefont
			{Shuang}}, \bibinfo {author} {\bibfnamefont {D.}~\bibnamefont {Ocampo}},
		\bibinfo {author} {\bibfnamefont {R.}~\bibnamefont {Namakian}}, \bibinfo
		{author} {\bibfnamefont {A.}~\bibnamefont {Ghasemi}}, \bibinfo {author}
		{\bibfnamefont {P.}~\bibnamefont {Dey}},\ and\ \bibinfo {author}
		{\bibfnamefont {W.}~\bibnamefont {Gao}},\ }\bibfield  {title} {\bibinfo
		{title} {{Kinetics of vacancy-assisted reversible phase transition in
				monolayer MoTe$_2$}},\ }\href {https://doi.org/10.1038/s43246-026-01078-0}
	{\bibfield  {journal} {\bibinfo  {journal} {Communications Materials}\
		}\textbf {\bibinfo {volume} {7}},\ \bibinfo {pages} {69} (\bibinfo {year}
		{2026})}\BibitemShut {NoStop}%
	\bibitem [{\citenamefont {Stach}\ \emph {et~al.}(2021)\citenamefont {Stach},
		\citenamefont {DeCost}, \citenamefont {Kusne}, \citenamefont
		{Hattrick-Simpers}, \citenamefont {Brown}, \citenamefont {Reyes},
		\citenamefont {Schrier}, \citenamefont {Billinge}, \citenamefont
		{Buonassisi}, \citenamefont {Foster}, \citenamefont {Gomes}, \citenamefont
		{Gregoire}, \citenamefont {Mehta}, \citenamefont {Montoya}, \citenamefont
		{Olivetti}, \citenamefont {Park}, \citenamefont {Rotenberg}, \citenamefont
		{Saikin}, \citenamefont {Smullin}, \citenamefont {Stanev},\ and\
		\citenamefont {Maruyama}}]{Stach_2021}%
	\BibitemOpen
	\bibfield  {author} {\bibinfo {author} {\bibfnamefont {E.}~\bibnamefont
			{Stach}}, \bibinfo {author} {\bibfnamefont {B.}~\bibnamefont {DeCost}},
		\bibinfo {author} {\bibfnamefont {A.~G.}\ \bibnamefont {Kusne}}, \bibinfo
		{author} {\bibfnamefont {J.}~\bibnamefont {Hattrick-Simpers}}, \bibinfo
		{author} {\bibfnamefont {K.~A.}\ \bibnamefont {Brown}}, \bibinfo {author}
		{\bibfnamefont {K.~G.}\ \bibnamefont {Reyes}}, \bibinfo {author}
		{\bibfnamefont {J.}~\bibnamefont {Schrier}}, \bibinfo {author} {\bibfnamefont
			{S.}~\bibnamefont {Billinge}}, \bibinfo {author} {\bibfnamefont
			{T.}~\bibnamefont {Buonassisi}}, \bibinfo {author} {\bibfnamefont
			{I.}~\bibnamefont {Foster}}, \bibinfo {author} {\bibfnamefont {C.~P.}\
			\bibnamefont {Gomes}}, \bibinfo {author} {\bibfnamefont {J.~M.}\ \bibnamefont
			{Gregoire}}, \bibinfo {author} {\bibfnamefont {A.}~\bibnamefont {Mehta}},
		\bibinfo {author} {\bibfnamefont {J.}~\bibnamefont {Montoya}}, \bibinfo
		{author} {\bibfnamefont {E.}~\bibnamefont {Olivetti}}, \bibinfo {author}
		{\bibfnamefont {C.}~\bibnamefont {Park}}, \bibinfo {author} {\bibfnamefont
			{E.}~\bibnamefont {Rotenberg}}, \bibinfo {author} {\bibfnamefont {S.~K.}\
			\bibnamefont {Saikin}}, \bibinfo {author} {\bibfnamefont {S.}~\bibnamefont
			{Smullin}}, \bibinfo {author} {\bibfnamefont {V.}~\bibnamefont {Stanev}},\
		and\ \bibinfo {author} {\bibfnamefont {B.}~\bibnamefont {Maruyama}},\
	}\bibfield  {title} {\bibinfo {title} {Autonomous experimentation systems for
			materials development: A community perspective},\ }\href
	{https://doi.org/https://doi.org/10.1016/j.matt.2021.06.036} {\bibfield
		{journal} {\bibinfo  {journal} {Matter}\ }\textbf {\bibinfo {volume} {4}},\
		\bibinfo {pages} {2702} (\bibinfo {year} {2021})}\BibitemShut {NoStop}%
	\bibitem [{\citenamefont {Kusne}\ \emph {et~al.}(2020)\citenamefont {Kusne},
		\citenamefont {Yu}, \citenamefont {Wu}, \citenamefont {Zhang}, \citenamefont
		{Hattrick-Simpers}, \citenamefont {DeCost}, \citenamefont {Sarker},
		\citenamefont {Oses}, \citenamefont {Toher}, \citenamefont {Curtarolo},
		\citenamefont {Davydov}, \citenamefont {Agarwal}, \citenamefont {Bendersky},
		\citenamefont {Li}, \citenamefont {Mehta},\ and\ \citenamefont
		{Takeuchi}}]{Kusne_2020}%
	\BibitemOpen
	\bibfield  {author} {\bibinfo {author} {\bibfnamefont {A.~G.}\ \bibnamefont
			{Kusne}}, \bibinfo {author} {\bibfnamefont {H.}~\bibnamefont {Yu}}, \bibinfo
		{author} {\bibfnamefont {C.}~\bibnamefont {Wu}}, \bibinfo {author}
		{\bibfnamefont {H.}~\bibnamefont {Zhang}}, \bibinfo {author} {\bibfnamefont
			{J.}~\bibnamefont {Hattrick-Simpers}}, \bibinfo {author} {\bibfnamefont
			{B.}~\bibnamefont {DeCost}}, \bibinfo {author} {\bibfnamefont
			{S.}~\bibnamefont {Sarker}}, \bibinfo {author} {\bibfnamefont
			{C.}~\bibnamefont {Oses}}, \bibinfo {author} {\bibfnamefont {C.}~\bibnamefont
			{Toher}}, \bibinfo {author} {\bibfnamefont {S.}~\bibnamefont {Curtarolo}},
		\bibinfo {author} {\bibfnamefont {A.~V.}\ \bibnamefont {Davydov}}, \bibinfo
		{author} {\bibfnamefont {R.}~\bibnamefont {Agarwal}}, \bibinfo {author}
		{\bibfnamefont {L.~A.}\ \bibnamefont {Bendersky}}, \bibinfo {author}
		{\bibfnamefont {M.}~\bibnamefont {Li}}, \bibinfo {author} {\bibfnamefont
			{A.}~\bibnamefont {Mehta}},\ and\ \bibinfo {author} {\bibfnamefont
			{I.}~\bibnamefont {Takeuchi}},\ }\bibfield  {title} {\bibinfo {title}
		{On-the-fly closed-loop materials discovery via bayesian active learning},\
	}\href {https://doi.org/10.1038/s41467-020-19597-w} {\bibfield  {journal}
		{\bibinfo  {journal} {Nature Communications}\ }\textbf {\bibinfo {volume}
			{11}},\ \bibinfo {pages} {5966} (\bibinfo {year} {2020})}\BibitemShut
	{NoStop}%
	\bibitem [{\citenamefont {Szymanski}\ \emph {et~al.}(2023)\citenamefont
		{Szymanski}, \citenamefont {Rendy}, \citenamefont {Fei}, \citenamefont
		{Kumar}, \citenamefont {He}, \citenamefont {Milsted}, \citenamefont
		{McDermott}, \citenamefont {Gallant}, \citenamefont {Cubuk}, \citenamefont
		{Merchant}, \citenamefont {Kim}, \citenamefont {Jain}, \citenamefont
		{Bartel}, \citenamefont {Persson}, \citenamefont {Zeng},\ and\ \citenamefont
		{Ceder}}]{Szymanski_2023}%
	\BibitemOpen
	\bibfield  {author} {\bibinfo {author} {\bibfnamefont {N.~J.}\ \bibnamefont
			{Szymanski}}, \bibinfo {author} {\bibfnamefont {B.}~\bibnamefont {Rendy}},
		\bibinfo {author} {\bibfnamefont {Y.}~\bibnamefont {Fei}}, \bibinfo {author}
		{\bibfnamefont {R.~E.}\ \bibnamefont {Kumar}}, \bibinfo {author}
		{\bibfnamefont {T.}~\bibnamefont {He}}, \bibinfo {author} {\bibfnamefont
			{D.}~\bibnamefont {Milsted}}, \bibinfo {author} {\bibfnamefont {M.~J.}\
			\bibnamefont {McDermott}}, \bibinfo {author} {\bibfnamefont {M.}~\bibnamefont
			{Gallant}}, \bibinfo {author} {\bibfnamefont {E.~D.}\ \bibnamefont {Cubuk}},
		\bibinfo {author} {\bibfnamefont {A.}~\bibnamefont {Merchant}}, \bibinfo
		{author} {\bibfnamefont {H.}~\bibnamefont {Kim}}, \bibinfo {author}
		{\bibfnamefont {A.}~\bibnamefont {Jain}}, \bibinfo {author} {\bibfnamefont
			{C.~J.}\ \bibnamefont {Bartel}}, \bibinfo {author} {\bibfnamefont
			{K.}~\bibnamefont {Persson}}, \bibinfo {author} {\bibfnamefont
			{Y.}~\bibnamefont {Zeng}},\ and\ \bibinfo {author} {\bibfnamefont
			{G.}~\bibnamefont {Ceder}},\ }\bibfield  {title} {\bibinfo {title} {An
			autonomous laboratory for the accelerated synthesis of inorganic materials},\
	}\href {https://doi.org/10.1038/s41586-023-06734-w} {\bibfield  {journal}
		{\bibinfo  {journal} {Nature}\ }\textbf {\bibinfo {volume} {624}},\ \bibinfo
		{pages} {86} (\bibinfo {year} {2023})}\BibitemShut {NoStop}%
	\bibitem [{\citenamefont {Sung}\ \emph {et~al.}(2024)\citenamefont {Sung},
		\citenamefont {Agarwal}, \citenamefont {El~Baggari}, \citenamefont {Kezer},
		\citenamefont {Goh}, \citenamefont {Schnitzer}, \citenamefont {Shen},
		\citenamefont {Chiang}, \citenamefont {Liu}, \citenamefont {Lu},
		\citenamefont {Sun}, \citenamefont {Kourkoutis}, \citenamefont {Heron},
		\citenamefont {Sun},\ and\ \citenamefont {Hovden}}]{Sung_2024}%
	\BibitemOpen
	\bibfield  {author} {\bibinfo {author} {\bibfnamefont {S.~H.}\ \bibnamefont
			{Sung}}, \bibinfo {author} {\bibfnamefont {N.}~\bibnamefont {Agarwal}},
		\bibinfo {author} {\bibfnamefont {I.}~\bibnamefont {El~Baggari}}, \bibinfo
		{author} {\bibfnamefont {P.}~\bibnamefont {Kezer}}, \bibinfo {author}
		{\bibfnamefont {Y.~M.}\ \bibnamefont {Goh}}, \bibinfo {author} {\bibfnamefont
			{N.}~\bibnamefont {Schnitzer}}, \bibinfo {author} {\bibfnamefont {J.~M.}\
			\bibnamefont {Shen}}, \bibinfo {author} {\bibfnamefont {T.}~\bibnamefont
			{Chiang}}, \bibinfo {author} {\bibfnamefont {Y.}~\bibnamefont {Liu}},
		\bibinfo {author} {\bibfnamefont {W.}~\bibnamefont {Lu}}, \bibinfo {author}
		{\bibfnamefont {Y.}~\bibnamefont {Sun}}, \bibinfo {author} {\bibfnamefont
			{L.~F.}\ \bibnamefont {Kourkoutis}}, \bibinfo {author} {\bibfnamefont
			{J.~T.}\ \bibnamefont {Heron}}, \bibinfo {author} {\bibfnamefont
			{K.}~\bibnamefont {Sun}},\ and\ \bibinfo {author} {\bibfnamefont
			{R.}~\bibnamefont {Hovden}},\ }\bibfield  {title} {\bibinfo {title}
		{{Endotaxial stabilization of 2D charge density waves with long-range
				order}},\ }\href {https://doi.org/10.1038/s41467-024-45711-3} {\bibfield
		{journal} {\bibinfo  {journal} {Nature Communications}\ }\textbf {\bibinfo
			{volume} {15}},\ \bibinfo {pages} {1403} (\bibinfo {year}
		{2024})}\BibitemShut {NoStop}%
	\bibitem [{\citenamefont {Li}\ \emph {et~al.}(2021)\citenamefont {Li},
		\citenamefont {Zaki}, \citenamefont {Garlea}, \citenamefont {Savici},
		\citenamefont {Fobes}, \citenamefont {Xu}, \citenamefont {Camino},
		\citenamefont {Petrovic}, \citenamefont {Gu}, \citenamefont {Johnson},
		\citenamefont {Tranquada},\ and\ \citenamefont {Zaliznyak}}]{Li_2021_2}%
	\BibitemOpen
	\bibfield  {author} {\bibinfo {author} {\bibfnamefont {Y.}~\bibnamefont
			{Li}}, \bibinfo {author} {\bibfnamefont {N.}~\bibnamefont {Zaki}}, \bibinfo
		{author} {\bibfnamefont {V.~O.}\ \bibnamefont {Garlea}}, \bibinfo {author}
		{\bibfnamefont {A.~T.}\ \bibnamefont {Savici}}, \bibinfo {author}
		{\bibfnamefont {D.}~\bibnamefont {Fobes}}, \bibinfo {author} {\bibfnamefont
			{Z.}~\bibnamefont {Xu}}, \bibinfo {author} {\bibfnamefont {F.}~\bibnamefont
			{Camino}}, \bibinfo {author} {\bibfnamefont {C.}~\bibnamefont {Petrovic}},
		\bibinfo {author} {\bibfnamefont {G.}~\bibnamefont {Gu}}, \bibinfo {author}
		{\bibfnamefont {P.~D.}\ \bibnamefont {Johnson}}, \bibinfo {author}
		{\bibfnamefont {J.~M.}\ \bibnamefont {Tranquada}},\ and\ \bibinfo {author}
		{\bibfnamefont {I.~A.}\ \bibnamefont {Zaliznyak}},\ }\bibfield  {title}
	{\bibinfo {title} {{Electronic properties of the bulk and surface states of
				Fe$_{1+y}$Te$_{1-x}$Se$_x$}},\ }\href
	{https://doi.org/10.1038/s41563-021-00984-7} {\bibfield  {journal} {\bibinfo
			{journal} {Nature Materials}\ }\textbf {\bibinfo {volume} {20}},\ \bibinfo
		{pages} {1221} (\bibinfo {year} {2021})}\BibitemShut {NoStop}%
\end{thebibliography}
\end{document}